\documentclass[nolinenumbers,trackchanges,twocolumn]{aastex701} 
\usepackage{verbatim}
\usepackage{siunitx}
\usepackage{enumitem} 
\usepackage{csquotes}
\usepackage{placeins}

\begin{document}

\title{The Impact of Simulation Resolution on Dwarf Galaxy \\ Stellar Stream Populations in FIRE}

\author[0009-0002-4295-9239]{Yasmine J. Meziani}
\affiliation{Division of Physics, Mathematics, and Astronomy, California Institute of Technology, Pasadena, CA 91125, USA} 
\email[show]{ymeziani@caltech.edu}

\author[0000-0001-5214-8822]{Nondh Panithanpaisal} 
\affiliation{Division of Physics, Mathematics, and Astronomy, California Institute of Technology, Pasadena, CA 91125, USA}
\affiliation{The Observatories of the Carnegie Institution for Science, 813 Santa Barbara Street, Pasadena, CA 91101, USA}
\email{}

\author[0000-0003-3939-3297]{Robyn E. Sanderson}
\affiliation{Department of Physics \& Astronomy, University of Pennsylvania, 209 S 33rd St, Philadelphia, PA 19104, USA}
\email{}

\author[0000-0003-0603-8942]{Andrew Wetzel} 
\affiliation{Department of Physics and Astronomy, University of California, Davis, CA 95616, USA}
\email{}

\author[0000-0003-0965-605X]{Pratik J. Gandhi}
\affiliation{Department of Astronomy, Yale University, New Haven, CT 06520, USA}
\email{pratik.gandhi@yale.edu}

\author[0000-0003-3729-1684]{Philip F. Hopkins}
\affiliation{Division of Physics, Mathematics, and Astronomy, California Institute of Technology, Pasadena, CA 91125, USA}
\email{}

\author[0000-0002-7352-6252]{Adriana Dropulic}
\affiliation{DARK, Niels Bohr Institute, University of Copenhagen, Jagtvej 155A, 2200 Copenhagen N, Denmark}
\email{}


\begin{abstract}
Cosmological simulations allow us to study galaxy-progenitor stream populations, allowing comparisons with observations to constrain galaxy formation physics. Previous work found that present-day galaxy-progenitor streams in FIRE have larger pericenters than observed in the MW, which might indicate that they tidally disrupt at larger distances, where the Galactic tides are weaker, than in the MW.
One possible explanation is that they tidally disrupt too quickly because of limited resolution. At lower resolution, low-mass dwarf galaxies can be puffier and more susceptible to disruption.
We build and compare dwarf-galaxy stream catalogs in two FIRE-2 cosmological zoom-in simulations with identical initial conditions (\texttt{m12i}) but different particle mass resolution ($7070 \, M_{\odot}$ and $880 \, M_{\odot}$). The higher-resolution run yields a larger dwarf galaxy stream population ($22$ streams versus $14$) using the same cutoff of 100 star particles, primarily because it resolves lower-mass streams (down to $~ 10^{5} \; \rm M_{\odot}$). Both simulations form streams at galactocentric radii $< 50 \; \rm kpc$, but the higher-resolution \textit{Triple Latte} simulation produces more streams at these small radii. We find that dwarf galaxies in the higher-resolution simulation begin disrupting at smaller pericenters than in the lower-resolution simulation. By cross-matching stream candidates, we identify counterparts for $\sim 89\%$ of the systems in the lower-resolution simulation. For matched systems that remain on similar orbits, the higher-resolution counterparts remain intact for longer. The orbital properties of streams in the higher-resolution simulation are broadly consistent with the MW dwarf-galaxy stream population, in contrast to the lower-resolution simulation. 

\end{abstract}

\keywords{\uat{Stellar streams}{2166}; \uat{Hydrodynamical simulations}{767}; \uat{Galaxy Structure}{622}}


\section{Introduction} \label{sec:intro}

The hierarchical nature of structure formation in a $\Lambda$CDM universe predicts that galaxies, such as the Milky Way (MW), form through the merger, accretion, and disruption of smaller systems over cosmic time. Evidence of this assembly history in the MW stellar halo is preserved through long-lived dynamical structures of past mergers and interactions \citep{johnston_fossil_1996}. As remnants of these accretion events, stellar streams offer a \enquote{fossil record} of the Galaxy's formation history and the properties of its dark matter halo \citep{bonaca_stellar_2024}.

MW stellar streams form through tidal disruptions of bound stellar systems, such as globular clusters or dwarf galaxies, as they interact with the host galaxy's gravitational potential. Stellar streams are used as complementary probes for investigating dark matter and galaxy formation in the local universe \citep{yoon_clumpy_2011,  carlberg_dark_2012, reino_galactic_2021, nibauer_constraining_2023, shipp_streams_2023, nibauer_galactic_2025}. Constraining the total number of accretion events onto the MW is important for testing galaxy formation models \citep{shipp_streams_2023}. Stellar streams are long-lived remnants of past accretion events that retain spatial and kinematic coherence, making them powerful tracers of a galaxy's formation history \citep{ibata_uncovering_2002}. Stellar streams originating from dwarf galaxies have also been used to constrain the enclosed dark matter distribution and halo shape of the MW \citep{springel_tidal_1999, dubinski_dynamical_1999, law_evidence_2009}. 

Cosmological simulations play a crucial role in the study of stellar streams by providing a theoretical framework to model their formation, evolution, and interaction with the host galaxy's dark matter halo \citep{panithanpaisal_galaxy_2021, shipp_streams_2023}. Recent studies, such as \citet{chiang_tidal_2025}, highlight how tidal forces can drive subhalos towards isotropic and cored structures, while \citet{chiang_universal_2026} explore numerical convergence criteria to model tidal stripping. By simulating the hierarchical assembly of galaxies and the accretion of satellite systems, these simulations predict the abundance, spatial distribution, and kinematic properties of stellar streams within galactic halos \citep{bullock_tracing_2005, helmi_stellar_2008, johnston_tracing_2008}. 

Currently, cosmological simulations predict that galaxies on scales similar to the MW contain a significant number of gravitationally bound subhalos within the virial radius of the host \citep{wetzel_reconciling_2016, panithanpaisal_galaxy_2021, barry_dark_2023}. Additionally, these simulations also show a complex network of tidal debris and disrupted satellites. The simulations further suggest that properties of these streams can vary depending on the galaxy's merger history, the shape of its dark matter halo, and the influence of the galactic disk and bulge \citep{bullock_tracing_2005, helmi_stellar_2008}. However, very few studies have been done of stellar stream populations in simulations at different numerical resolutions, which are important for understanding how well streams can be detected and how their properties compare with observed streams \citep{panithanpaisal_galaxy_2021, shipp_streams_2023, riley_auriga_2025, shipp_auriga_2025, riley_auriga_2026}. Understanding these resolution effects is important because numerical limitations can change the predicted stream population and may be mistaken for physical differences between simulations and observations. For example, \citet{shipp_streams_2023} found that dwarf-galaxy streams in FIRE have larger pericenters than those observed in the MW, but it is unclear whether this difference is physical or partly caused by limited numerical resolution.

Previous resolution tests in cosmological simulations have shown that the ability to form, track, and characterize stellar streams is sensitive to numerical resolution \citep[e.g.,][]{riley_auriga_2025}. Earlier work demonstrated that at low resolution, dark matter subhalos get stripped of mass artificially quickly and cannot be tracked below the resolution limit \citep[e.g.,][]{van_den_bosch_dark_2018, errani_asymptotic_2021}. 

A recent paper analyzing the Auriga cosmological zoom-in simulations of MW-mass galaxies found that the stellar morphologies and disruption states of accreted satellites, including stellar streams, are similar across different resolutions, with higher resolution revealing marginally surviving systems rather than qualitatively new structures \citep{riley_auriga_2025}. However, there has not been a dedicated resolution test focusing on quantifying how many additional streams appear at higher resolution or how individual streams compare across different resolution FIRE simulations. 


Over the past two decades, wide-field surveys such as SDSS, Gaia, and DES have revealed dozens of stellar streams in the MW halo, sparking interest in using them as probes of both galactic dynamics and fundamental physics \citep{bonaca_stellar_2024}. Deeper surveys have always uncovered new stellar streams (e.g., 
\citealt{koposov_discovery_2014, martin_pandas_2014,jethwa_discovery_2018, shipp_stellar_2018, li_southern_2019, martinez-delgado_hidden_2023, martinez-delgado_stellar_2025}); however, it is more difficult for surveys to detect streams further out in the halo because of lower surface brightness and lower proper motions. Thus, cosmological simulations paired with observations are critical to creating a full map of stellar streams in our MW.

In this paper, we create and compare stream catalogs for two separate FIRE-2 zoom-in simulations of the same MW-mass halo, \texttt{m12i}, at different resolutions.
The first version of m12i is part of the standard FIRE-2 \textit{Latte} suite \citep[introduced in][]{wetzel_reconciling_2016}, while the second is a resimulation of m12i at $8 \times$ higher mass resolution and is the first of the FIRE-2 \textit{Triple Latte} suite (Wetzel et al., in prep.).

Section \ref{sec:simulations} goes over the details of the simulations used. For each simulation we build a stream catalog, identifying and classifying stream progenitors based on the selection criteria defined in Section \ref{sec:stream selection}. In Section \ref{sec:results}, we compare the global properties of the stream catalogs, including their radial, mass, eccentricity, and orbital tilt distributions, as well as stream pericenters and apocenters. We also compare individual properties of a subset of cross-matched streams across the two simulations in Section \ref{sec:matching}. Section \ref{sec:discussions} discusses convergence of our catalog across resolution levels, catalog completeness, and comparisons between FIRE and Milky Way orbital properties. Section \ref{sec:conclusions} summarizes the main conclusions.

\section{Simulations}
\label{sec:simulations}

Cosmological baryonic zoom-in simulations of galaxy formation achieve resolutions that allow for the study of dwarf galaxies around MW-mass hosts. In this work, we focus on the MW-mass simulations of the \texttt{m12i} halo from the FIRE-2 \textit{Latte} suite introduced in \citep{wetzel_reconciling_2016} and the \textit{Triple Latte} suite (to be introduced in Wetzel et~al.~in prep).
These are cosmological zoom-in baryonic simulations of MW-mass galaxies from the Feedback in Realistic Environments (FIRE) project \citep{hopkins_galaxies_2014}, using the FIRE-2 model \citep{hopkins_fire-2_2018}.
Galaxies in the \textit{Latte} suite have (initial) mass resolution of $7070\; M_\odot$ for gas and stars and $35,000 \; M_\odot$ for dark matter,
with (Plummer-equivalent) gravitational force softenings of 4 pc for stars and 40 pc for dark matter.
The \textit{Triple Latte} suite has $8 \times$ higher mass resolution and $2 \times$ better spatial resolution. Specifically, the (initial) mass resolution is $880\; M_\odot$ for gas and stars and $4400 \; M_\odot$ for dark matter, with (Plummer-equivalent) gravitational force softenings of 2 pc for stars and 20 pc for dark matter.
In all cases, the gas softening is adaptive, to match the hydrodynamic smoothing, down to a minimum of 1 pc.

Among the 8 MW-mass halos in the FIRE-2 \textit{Latte} suite, \texttt{m12i} has a relatively high number of stellar streams, providing better statistics. The $7070 M_\odot$ resolution version of m12i -- with its full snapshot data as well as halo catalogs and merger tree -- is included in the FIRE-2 public data release \citep{wetzel_public_2023, wetzel_second_2025}. Henceforth in this work, we use \textit{Latte} and \textit{Triple Latte} to refer to the versions of \texttt{m12i}, specifically.

We generated these simulations with \texttt{GIZMO} \citep{hopkins_new_2015}, which uses a mesh-free, finite-mass Lagrangian Godunov method for the hydrodynamics solver and a TREE+PM gravity solver optimized from \texttt{GADGET-3} \citep{springel_gadget_2001}. 
Importantly, the simulations resolve the clumpy, multi-phase inter-stellar medium in galaxies and leverage physically motivated heating and cooling processes for gas across $10-10^{10}$ K in temperature and accounting for $11$ element species.
Star formation and stellar feedback are governed by the FIRE-2 model \citep{hopkins_fire-2_2018}. Stars form in locally self-gravitating, molecular gas above a density threshold of $n_{\rm H} > 1000 \;{\rm cm}^{-3}$, and once eligible for star formation, a gas cell converts to a star particle on a local free-fall time \citep{hopkins_general_2013}. Each star particle represents a simple stellar population following a \citet{kroupa_variation_2001} initial mass function. FIRE-2 evolves star particles based on stellar population models from \textsc{Starburst99} \citep{leitherer_starburst99_1999}. Stellar feedback prescriptions include radiation pressure, photoionization, stellar winds, and core-collapse and Type Ia supernovae, with injection rates and nucleosynthetic yields derived from \citet{iwamoto_nucleosynthesis_1999, leitherer_starburst99_1999, mannucci_two_2006, nomoto_nucleosynthesis_2006, wiersma_effect_2009}. Additionally, the simulations account for the global effects of cosmic reionization by using a spatially uniform, redshift-dependent UV/X-ray background \citep{faucher-giguere_new_2009}.

In this work, we adopt the following terminology. We define a halo as a gravitationally bound dark matter structure and a subhalo as a halo that resides within the virial radius of a more massive host halo. For each halo/subhalo, we define its virial radius $R_{200\rm m}$ as the radius enclosing $200\times$ the mean matter density of the Universe, and its virial mass $M_{200 \rm m}$ as the mass enclosed inside $R_{200\rm m}$. We define a galaxy as the stellar component associated with a halo or subhalo and a satellite galaxy as a galaxy within a subhalo orbiting a more massive host. We define a progenitor as the original galaxy (pre-infall into the host halo) that gives rise to a stream or disrupted debris. We classify systems as intact (or bound) if they remain gravitationally self-bound at $z=0$. A subhalo is considered unbound if \texttt{Rockstar} can no longer identify the (sub)halo as a coherent structure in phase space. We classify stellar debris as a stream if it remains spatially coherent and as phase-mixed if it has dispersed such that no coherent structure is visible.

The FIRE-2 simulations generate MW-mass galaxies that broadly resemble observed properties of the MW. These simulated galaxies exhibit thin stellar disks with features such as spiral arms and central bars \citep{ma_structure_2017, debattista_formation_2019, hopkins_fire-2_2018, sanderson_synthetic_2020, bellardini_3d_2021, bellardini_3d_2022, yu_born_2023, mccluskey_disc_2024, ansar_bar_2025}, realistic populations of giant molecular clouds \citep{guszejnov_evolution_2020, benincasa_live_2020}, stellar halos from disrupted satellite galaxies \citep{sanderson_reconciling_2018, horta_observable_2023}, and satellite dwarf galaxies that are not affected by the "missing satellites" or "too-big-to-fail" problems \citep{wetzel_reconciling_2016, garrison-kimmel_star_2019, garrison-kimmel_local_2019, samuel_profile_2020, samuel_extinguishing_2022}.
Additionally, the population of satellite galaxies around each host resembles the MW’s in key aspects, including their stellar mass function, star formation histories, size–mass relation, and internal velocity dispersion \citep{wetzel_reconciling_2016, garrison-kimmel_star_2019, garrison-kimmel_local_2019, santistevan_orbital_2023}. The simulations also naturally produce stellar halos from disrupted satellites and match the observed radial distribution of surviving satellites, particularly within the inner $\sim 150\; \mathrm{kpc}$ \citep{samuel_profile_2020, horta_observable_2023}. A key factor behind these consistencies is that FIRE-2 forms massive central galaxies that are capable of tidally disrupting low-mass subhalos via gravitational tidal forces, leading to fewer overly dense satellites \citep{garrison-kimmel_not_2017}. While globular cluster-like streams do not form due to resolution limits, the simulations still resolve disrupted dwarf galaxies sufficiently to produce a realistic population of tidal streams from satellite galaxies \citep{shipp_streams_2023}.


Dark matter halos and subhalos in each snapshot of the simulations are identified and processed using \texttt{Rockstar} \citep{behroozi_rockstar_2013} to produce halo-catalogs. The resulting catalogs are then linked across time with \texttt{consistent-trees} \citep{behroozi_gravitationally_2013} to construct merger trees. After the merger tree is constructed, star particles are assigned to each halo and subhalo identified by \texttt{Rockstar} in each snapshot. This is done by selecting star particles within $0.8\times$ the halo's/subhalo's virial radius and with velocities less than twice the halo circular velocity relative to the halo center (for additional details, see \citet{necib_under_2019, samuel_profile_2020, gandhi_testing_2024}). However, the halo catalog-based star particle assignment fails to capture nearly all stream stars. Thus we use these assignments as a starting point to collect the full stellar distribution associated with the progenitor galaxy of each stream. All postprocessing is performed with the \texttt{gizmo\_analysis} and \texttt{halo\_analysis} packages \citep{wetzel_gizmoanalysis_2020, wetzel_haloanalysis_2020}.

\section{Stream Selection}
\label{sec:stream selection}

Streams for each simulation were selected based on a specific set of criteria. Here we explain the steps used to identify the streams and their progenitors. Methods outlined here are mostly consistent with those mentioned in \citet{panithanpaisal_galaxy_2021}.

\subsection{Stream Candidate Identification}\label{subsec:stream candidate identification}
Starting from the star particles assigned to each subhalo in the merger tree, we select subhalos that contain nonzero stellar mass and were bound at some point in the past, regardless of whether they remain bound at the present day. Specifically, we consider bound subhalos between $z \sim 0.0 - 1.0$ with at least $100$ star particles that lie within the virial radius of the host at the time it is identified. This $100$-particle cut corresponds to an approximate lower stellar mass limit of $M_{\star} \sim 6.5 \times 10^{5} \; M_{\odot}$ in \textit{Latte} and $M_{\star} \sim 6.2 \times 10^{4} \; M_{\odot}$ in \textit{Triple Latte}, accounting for stellar mass loss. Therefore, all analysis in this paper applies only to intact satellites, streams, and phase-mixed systems above these approximate stellar mass limits.
We choose an extended time window because at the present day, most stream progenitors have already become unbound, and thus are not identified by the halo finder \texttt{Rockstar} \citep{behroozi_gravitationally_2013, behroozi_rockstar_2013}. We limit the selection window to $z \leq 1$ because subhalos that become unbound earlier are expected to contain satellites that are largely phase-mixed by the present day \citep{panithanpaisal_galaxy_2021}. 

Each substructure identified is tracked back in time using the halo merger tree to when it was first identified by \texttt{Rockstar}, 
and forward to the present day if it remains bound, or until it is no longer identified in the merger tree.
We first identify all the star particles that ever belonged to the subhalo at any point in time.



These star particles are then tracked to the present day to obtain their positions. To ensure that we have gathered all the star particles belonging to a specific subhalo, we collect all star particles that ever belonged to it. To do this, we use the merger tree to gather star particle indices in intervals of five snapshots while the subhalos are identified and tracked. We then take the union of all the star particles, ensuring no particle indices are double counted. We follow this procedure for all stream candidates identified in both \texttt{m12i} \textit{Latte} and \textit{Triple Latte} simulations.

Once we have a set of accreted stars, we check for contamination, which we define as star particles that are not physically associated with the subhalo but were misassigned to it by the halo finder. This misassignment generally occurs near pericenter or during disk crossings, where the stellar density contrast between the subhalo and the background star particles is low. We use each star particle’s formation time to determine whether it formed within the virial radius of the subhalo. We remove any star particles that were outside the subhalo’s virial radius at their formation time.



\subsection{Stream Candidate Classification}\label{subsec:stream candidate classification}




\begin{table*}[t]
\centering
\caption{
Summary of stream, phase-mixed, and intact satellite classifications at the present day for each host.
Column~(1) lists the host identifier.
Columns~(2) and (3) give the initial baryonic and dark-matter particle masses, respectively.
Columns~(4) and (5) give the Plummer-equivalent gravitational force softenings for stars and dark matter, respectively.
Column~(6) gives the approximate minimum stellar mass at $z=0$ corresponding to the adopted cutoff of 100 star particles.
Column~(7) gives the virial mass of the host at $z=0$.
Columns~(8)--(11) report the number of systems classified as streams, phase-mixed, and intact satellites, respectively.
}



\label{tab:host_summary}

\setlength{\tabcolsep}{4pt}

\begin{tabular}{lcccccccccc}
\hline
Host &
\shortstack{$m_{\rm baryon}$\\$[M_\odot]$} &
\shortstack{$m_{\rm DM}$\\$[M_\odot]$} &
\shortstack{$\epsilon_\star$\\$[\mathrm{pc}]$} &
\shortstack{$\epsilon_{\rm DM}$\\$[\mathrm{pc}]$} &
\shortstack{$M_{\star,\min}$\\$[M_\odot]$} &
\shortstack{$M_{\rm host}(z=0)$\\$[M_\odot]$} &
\shortstack{$N_{\rm stream}$\\total} &
\shortstack{$N_{\rm stream}$\\above \textit{Latte}\\mass resolution} &
$N_{\rm phase\text{-}mixed}$ &
$N_{\rm intact}$ \\
\hline
\texttt{m12i} \textit{Triple Latte}
& $880$
& $4400$
& $2$
& $20$
& $\sim6.2 \times 10^4$
& $1.07 \times 10^{12}$
& 22
& 16
& 14
& 9 \\
\texttt{m12i} \textit{Latte}
& $7070$
& $35{,}000$
& $4$
& $40$
& $\sim 6.5 \times 10^5$
& $1.05 \times 10^{12}$
& 14
& --
& 2
& 3 \\
\hline
\end{tabular}
\end{table*}

To determine whether an accreted satellite galaxy is classified as a stream, intact satellite, or phase-mixed structure at the present day, we first divide subhalos into two categories: those identified by the merger tree at $z=0$, and those that are not.

Subhalos identified by the merger tree at $z=0$ are classified as either intact subhalos or streams. The merger tree readily identifies intact subhalos, as well as streams whose progenitors are still at least partially bound. To distinguish between these two categories, we use two quantities: the median pairwise distance between star particles, $d_{\rm med}$, and the fraction of star particles that remain gravitationally bound to the subhalo ($f_{\rm bound}$).


For a subhalo containing $N_\star$ star particles at positions $\{\mathbf{x}_i\}_{i=1}^{N_\star}$, we compute all unique pairwise separations
\begin{equation}
d_{ij} = \left| \mathbf{x}_i - \mathbf{x}_j \right|, \qquad 1 \le i < j \le N_\star,
\end{equation}
and we define the median pairwise distance as
\begin{equation}
d_{\rm med} \equiv \mathrm{median}\!\left(\left\{ d_{ij} \right\}_{i<j}\right).
\end{equation}

Intact satellite galaxies are compact in position space, and therefore have relatively small values of $d_{\rm med}$ ($d_{ij}  \lesssim 2R_\textrm{vir,sub}$), where $R_{\rm vir,sub}$ is the subhalo’s own virial radius. On the other hand, stars in a stream are spread across the stellar halo of the host galaxy. For each dwarf galaxy, we compute $d_{\rm med}$ using all star particles and compare it to a threshold defined as a fixed fraction of the host virial radius. We adopt a threshold of $\frac{1}{2}R_{\rm vir}$ for both \textit{Latte} and \textit{Triple Latte}. Subhalos with $d_{\rm med} > R_{\rm vir}/2$ are classified as streams, while those with $d_{\rm med} < R_{\rm vir}/2$ are classified as intact. For subhalos containing more than $10{,}000$ star particles, we classify them by visual inspection to avoid the computational cost of evaluating all pairwise separations.

The bound fraction is defined as
\begin{equation}
f_{\rm bound} \equiv \frac{N_{\star,\mathrm{bound}}}{N_{\star, \rm total}},
\end{equation}
where $N_{\star,\mathrm{bound}}$ is the number of star particles that remain gravitationally bound to the subhalo at $z=0$, and $N_{\star, \rm total}$ is the total number of unique star particles that ever belonged to the subhalo at any time of its history. 
We use the merger tree to identify the currently bound stellar particles, since it tracks only the gravitationally bound components of a subhalo. We require that at least $90\%$ of the stellar mass remain bound for a system to be classified as intact. If both methods produce the same classification, we adopt that classification for the subhalo. If the two methods produce different classifications, we classify the subhalo by visual inspection.

Subhalos not identified by the merger tree at $z=0$ are classified as either streams or phase-mixed structures. For subhalos with more than $10^4$ star particles ($\rm N_{\star} > 10^{4}$), we use visual classification exclusively, as applying the full diagnostic becomes computationally time consuming and the number of such systems is small. For streams with $\rm N_{\star} < 10^{4}$ star particles, we use a quantitative diagnostic based on the local velocity dispersion of star particles within each galaxy. Specifically, we compare the local velocity dispersion of the stellar debris to the benchmark used in \citet{panithanpaisal_galaxy_2021} to guide the classification. In most cases, this diagnostic produces the same classification as visual inspection; however, we reclassify seven objects based on visual inspection when the morphology clearly indicates a different category. We note that the benchmark in \citet{panithanpaisal_galaxy_2021} is based on the \textit{Latte} simulation suite at $7070 M_\odot$ baryonic mass resolution, and will need to be recalibrated for \texttt{m12i} \textit{Triple Latte} at $880 \; \rm M_\odot$ resolution. Therefore, we use this diagnostic as a guideline rather than a strict threshold. 
The resulting classification counts for each host are summarized in Table \ref{tab:host_summary}.

\section{Population Analysis} \label{sec:results}

\subsection{Population Comparisons}\label{subsec:population statistics}

\begin{figure}
    \centering
    \includegraphics[width=\columnwidth]{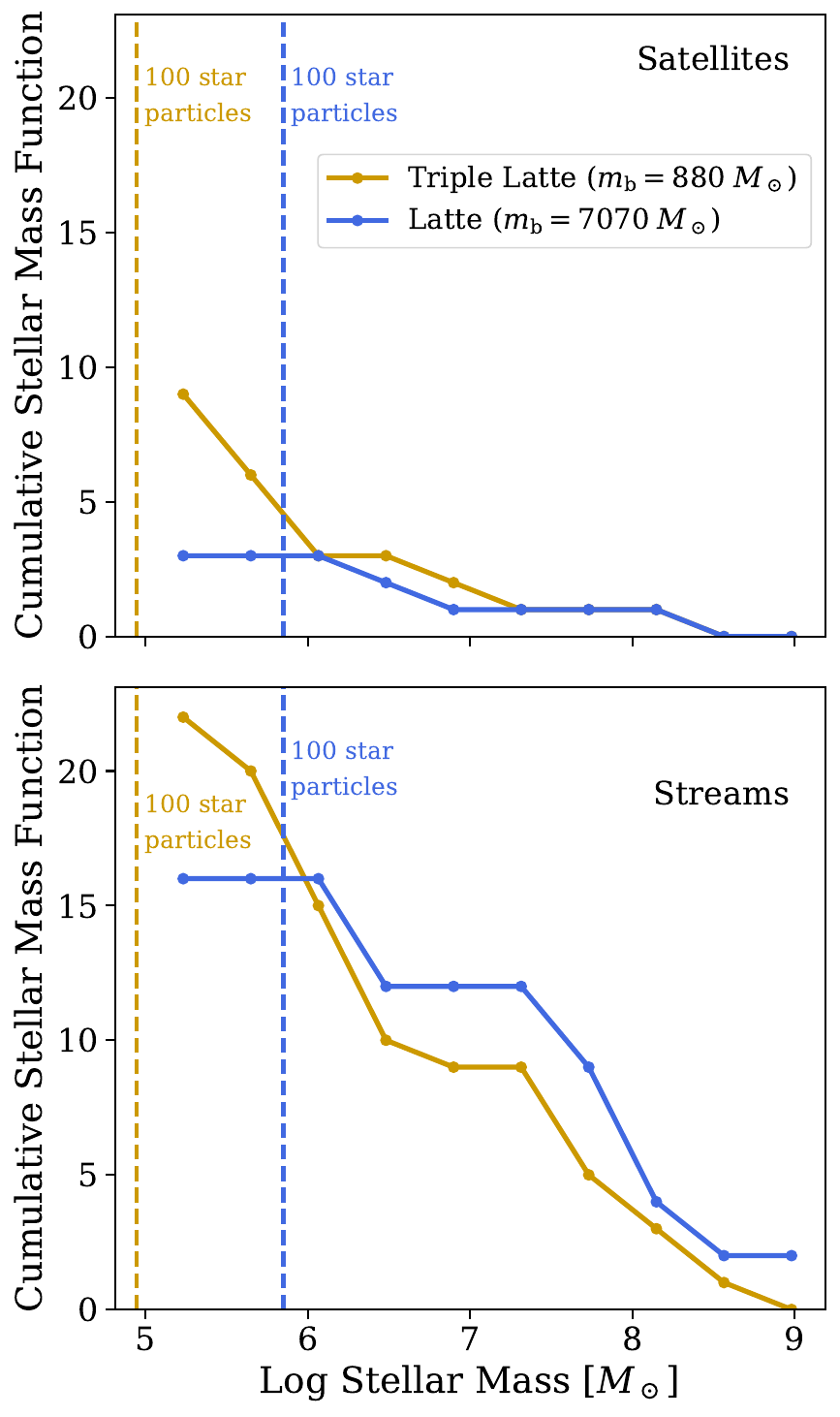}
    \caption{
    Stellar mass cumulative distribution plots for satellites (top) and streams (bottom) at present day in \texttt{m12i} \textit{Latte} (blue) and \textit{Triple Latte} (gold). We find more satellites and streams in \textit{Triple Latte} compared to \textit{Latte}. We only find $3$ satellites in \textit{Latte} as opposed to $9$ satellites in \citet{wetzel_reconciling_2016} because some of the systems classified as satellites in that work are classified as streams in this paper.
    }
    \label{fig:mass distribution}
\end{figure}

\begin{figure}
    \centering
    \includegraphics[width=\columnwidth]{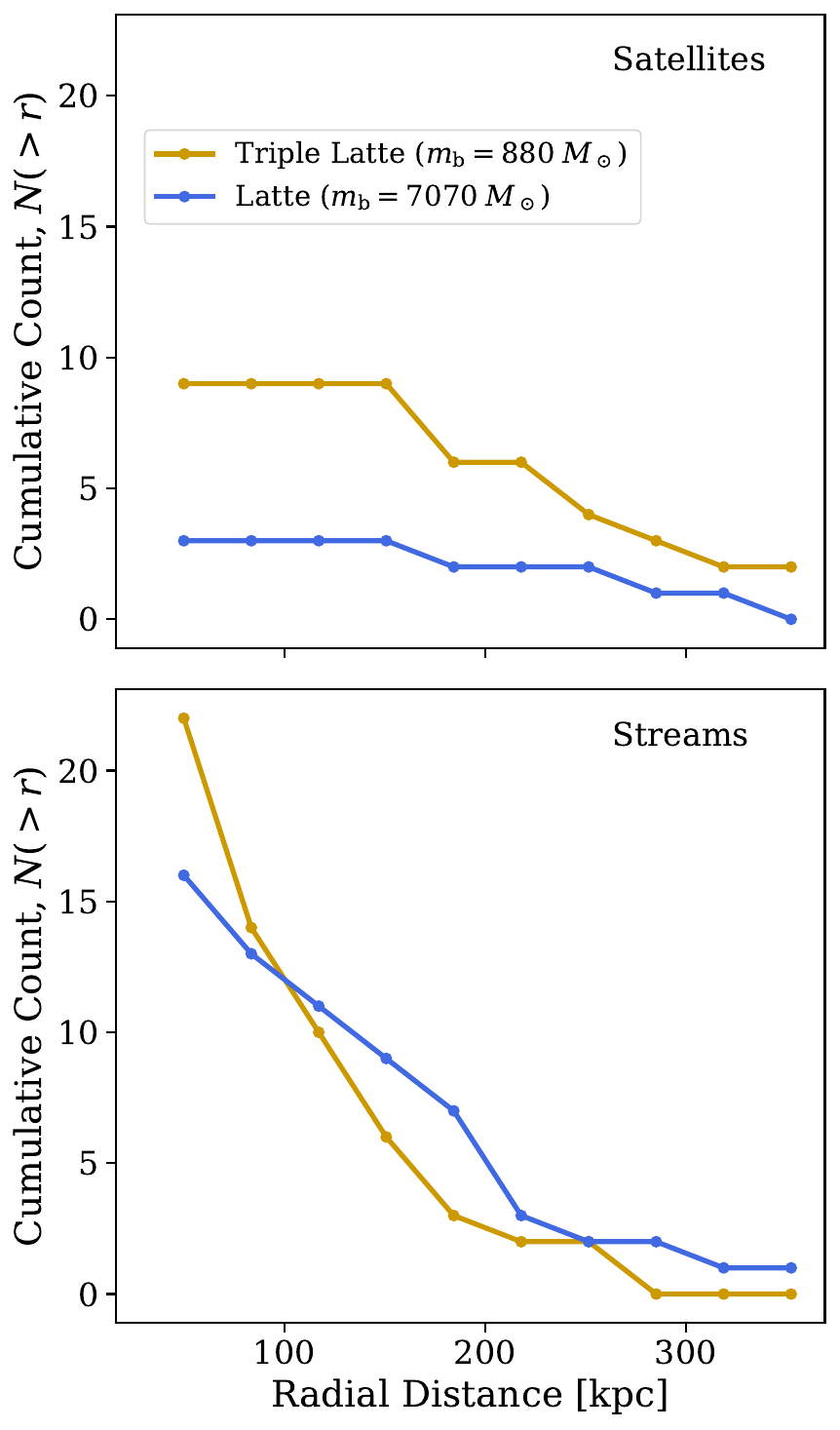}
    \caption{Cumulative radial distribution plots of streams (bottom) and satellites (top) at present day in \texttt{m12i} \textit{Latte} (blue) and \textit{Triple Latte} (gold). Color notation is as in Figure \ref{fig:mass distribution}. The 100-star-particle threshold (see Section \ref{subsec:stream candidate identification}) corresponds to $M_{\star}\sim6.5\times10^{5}\,M_{\odot}$ in \textit{Latte} and $\sim6.2\times10^{4}\,M_{\odot}$ in \textit{Triple Latte}. \textit{Triple Latte} has more streams at small radii.}
    \label{fig:radial distribution}
\end{figure}

\begin{figure}
    \centering
    \includegraphics[width=\columnwidth]{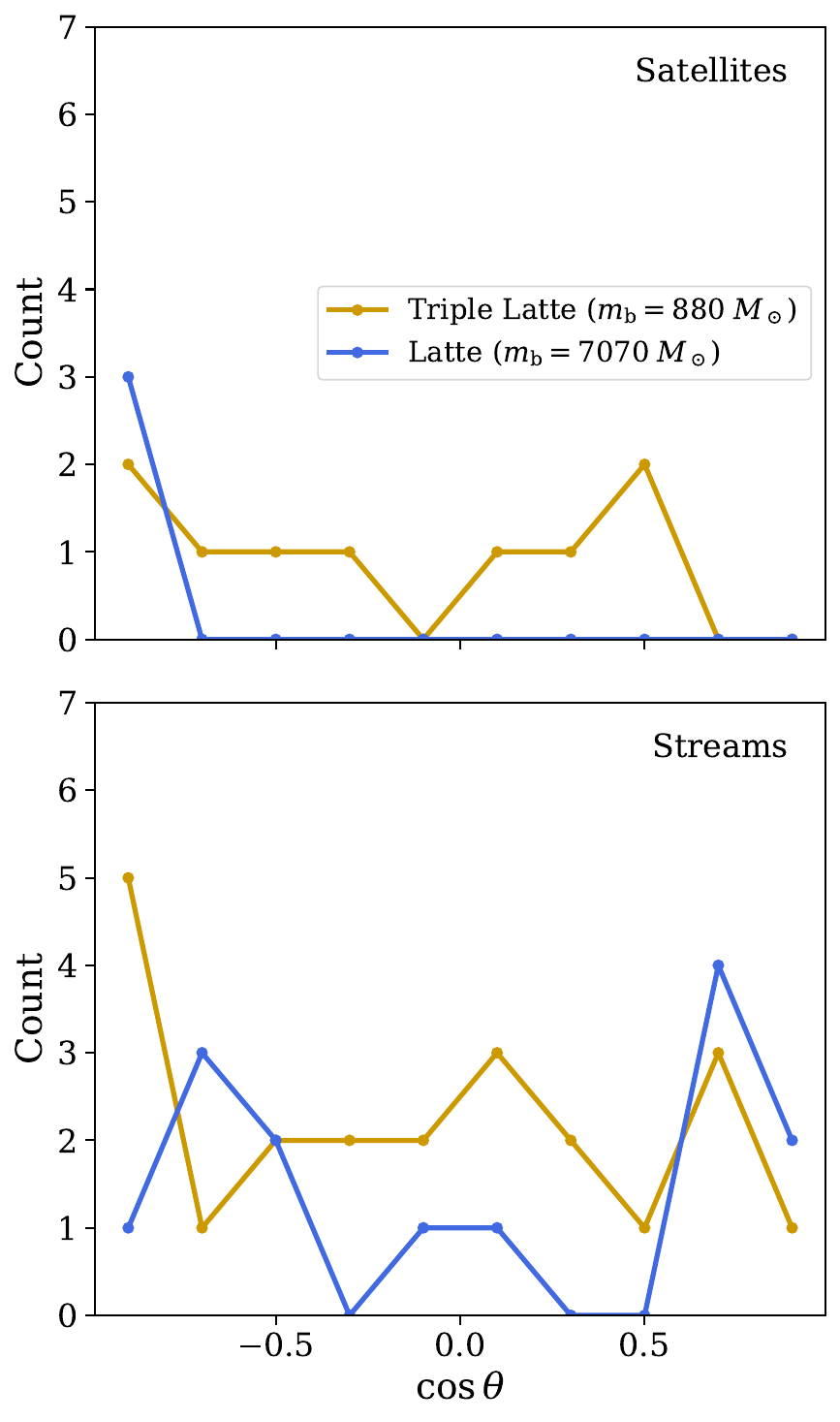}
    \caption{Distribution of orbital tilt at the present day for satellites and at the time of stream formation for streams. Orbital tilt is defined as the angle $\theta$ between a subhalo’s orbital angular momentum and the disk angular momentum. We use the same color assignment as in Figure~\ref{fig:mass distribution}. Here, $\cos(\theta)=1$ corresponds to prograde motion, $\cos(\theta)=-1$ to retrograde motion, and $\cos(\theta)=0$ to orbits perpendicular to the disk. In \texttt{m12i} \textit{Triple Latte} (gold), streams (bottom) show no clear preference for prograde or retrograde orbits, while satellites (top) favor retrograde orbits, although this trend is not statistically significant and a larger set of simulations is needed to test its robustness. In \textit{Latte} (blue), intact satellites also favor retrograde motion, while streams show no clear preference.}
    \label{fig:eccentricity and orbital tilt}
\end{figure}

\begin{figure}[t!] 
    \centering 
    \includegraphics[width=\columnwidth]{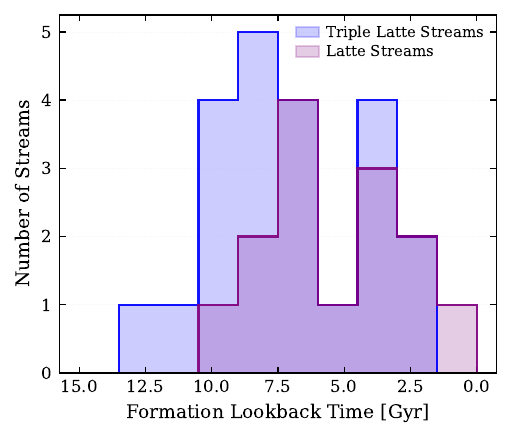} 
    \caption{Distribution of stream formation times for the \textit{Latte} and \textit{Triple Latte} streams. The \textit{Triple Latte} distribution shows a larger number of streams forming at earlier lookback times, while the \textit{Latte} distribution is shifted somewhat toward more recent stream formation times. Although the two distributions overlap substantially, the \textit{Latte} streams appear to form more recently on average than the \textit{Triple Latte} streams.}
    \label{fig:disruption distribution}
\end{figure}

In total, we identify $19$ stream candidates in \texttt{m12i} \textit{Latte} and $45$ candidates in \textit{Triple Latte} above our stellar mass threshold of $\sim 6.2 \times 10^4 \; \rm M_{\odot}$.
The larger number of objects in \textit{Triple Latte} reflects its higher mass resolution. In \textit{Triple Latte}, the minimum resolved stellar mass for our $100$ star particle threshold is $M_\star \sim 6.2 \times 10^4 \; \mathrm{M}{\odot}$, compared to $M\star \sim 6.5 \times 10^5 \; \mathrm{M}_{\odot}$ in \textit{Latte}. This lower mass limit for \textit{Triple Latte} allows us to resolve lower-mass subhalos and satellite galaxies. For \textit{Latte}, this sample consists of $14$ streams, $3$ intact satellites, and $2$ phase-mixed systems. In \textit{Triple Latte} we find $22$ streams, $9$ intact satellites, and $14$ phase-mixed systems (see Table \ref{tab:host_summary}). We identify $3$ intact satellites in \textit{Latte}, compared to $9$ in \citet{wetzel_reconciling_2016}, because some systems classified as satellites in that work are classified as streams in this paper.

For both \textit{Latte} and \textit{Triple Latte}, we characterize the individual distributions of several properties derived for each stream and satellite: stellar mass, distance from host, 
orbital tilt, and eccentricity. To ensure a clear comparison between intact satellites and streams, we exclude phase-mixed systems from the analysis in this section. We do not include the phase-mixed systems because our catalog for phase-mixed debris is highly incomplete by design, since we only consider bound subhalos between $z \sim 0.0-1.0$ (see Section \ref{subsec:stream candidate identification}). Many phase-mixed systems have already become gravitationally unbound by $z=1$.

Figure \ref{fig:mass distribution} displays the cumulative stellar mass distribution for both streams and intact satellites at present day for both \textit{Latte} and \textit{Triple Latte} simulations. The stellar mass was evaluated at the present day using the star particles gathered using the methods outlined in Section \ref{subsec:stream candidate identification}; vertical lines mark the mass corresponding to $100$ star particles.

Compared with the findings of \citet{panithanpaisal_galaxy_2021}, we find that the total numbers of streams and intact satellites depend on simulation resolution, primarily through the lower-mass systems resolved in \textit{Triple Latte}. At a fixed stellar mass threshold corresponding to the \textit{Latte} resolution limit, however, the number of streams is similar in the two simulations, with $16$ streams in \textit{Triple Latte} compared to $14$ streams in \textit{Latte}. In \textit{Latte}, the small number of intact satellites and the mass resolution limit make it difficult to identify trends with stellar mass. Low-mass stream progenitors are not as well resolved in \textit{Latte} as in \textit{Triple Latte}, contributing to the larger number of streams and satellites at lower masses in the cumulative distributions shown in Figure~\ref{fig:mass distribution}.
In contrast, \textit{Triple Latte} better resolves stream progenitors and intact satellites at lower stellar masses ($\lesssim 10^{6} \; \mathrm{M}_{\odot}$). This difference is expected from our $100$-star-particle selection criteria, which corresponds to a higher stellar-mass limit in \textit{Latte}. As a result, systems below $\sim 10^{6} \; \rm M_{\odot}$ are not included in the \textit{Latte} catalog.
A complementary analysis (Gandhi et al., in prep) will discuss the distribution of intact satellites in Triple Latte down to lower stellar masses than in this study.

Figure \ref{fig:radial distribution} shows the present-day cumulative galactocentric radial distribution of streams and satellites. For both \textit{Latte} and \textit{Triple Latte}, stellar streams are located at smaller radii ($< 100$ kpc) on average than intact satellites. However, this trend is more pronounced in the higher-resolution \textit{Triple Latte} simulation, which has a higher concentration of streams within $50$ kpc from the host than \textit{Latte}. This is expected, since progenitors on orbits with smaller galactocentric radii are more likely to be tidally disrupted. The higher-resolution simulation also tracks the bound remnants of stream progenitors deeper into the host halo. Stream progenitors on average were accreted over $8$ Gyr ago, indicating that longer exposure to the host's environment increases the likelihood of stream formation (see Figure \ref{fig:peri and apo}). 

Figure \ref{fig:eccentricity and orbital tilt} shows the distribution of orbital tilt values for intact satellites and streams. To calculate the orbital tilt, we first compute the orbital angular momentum per unit mass of each bound object relative to the host center at the present day for intact satellites and at the time of stream formation for streams, then we compute the angle $\theta$ between this orbital angular momentum and the disk's angular momentum. Thus, subhalos that orbit in the disk plane have $\cos(\theta) = \pm1$, while subhalos that have orbits perpendicular to the disk plane have $\cos(\theta) = 0$. Prograde orbits correspond to a distribution of $\cos(\theta) = +1$ while retrograde orbits correspond to $\cos(\theta) = -1$. 

The distribution of $\cos(\theta)$ for streams and intact satellites within the virial radius across both \textit{Latte} and \textit{Triple Latte} is shown in Figure \ref{fig:eccentricity and orbital tilt}. 
Previous studies suggest that streams may favor $\cos(\theta)=1$, while intact satellites may favor $\cos(\theta)=-1$ \citep{donghia_resonant_2009}. The stream distributions show no clear preference for prograde or retrograde orbits in either simulation, while intact satellites favor retrograde orbits with $\cos(\theta)\sim -1$ in both \textit{Latte} and \textit{Triple Latte}.
However, the differences in orbital tilt distributions may reflect small-number statistics, or host-to-host scatter. A larger sample of simulations would be needed to make robust claims about the orbital tilt distributions of streams and intact satellites.

This behavior may be related to interactions between satellites and non-axisymmetric structure in the stellar disk. For a perfectly axisymmetric disk, whether a satellite is on a prograde or retrograde orbit would not affect the strength of its tidal interaction with the disk. However, non-axisymmetric features can potentially produce stronger or more coherent tidal interactions for satellites on prograde orbits, increasing the likelihood that they are stripped into streams \citep{donghia_resonant_2009}. This differs from the results presented in \citet{panithanpaisal_galaxy_2021}; however, their analysis differs from ours in that it considers a cumulative distribution across $13$ different MW-mass galaxies from FIRE-2, whereas here we focus on a single host galaxy. Indeed, \citet{panithanpaisal_galaxy_2021} note substantial host-to-host scatter in the orbital tilt distribution, where orbital tilt measures the angle between a subhalo's orbital angular momentum and the host disk angular momentum. 

\citet{santistevan_modelling_2024} found no difference in the subhalo orbital evolution depending on the disk orientation in idealized tests. However, these tests used an idealized setup with a fixed background potential, which may not capture the full range of effects present in cosmological simulations. Therefore, disk orientation alone is unlikely to explain the differences we find in our orbital tilts. Instead, host-to-host scatter and differences in the cosmological assembly and orbital histories of the satellite populations may contribute to the orbital tilt distribution seen in Figure \ref{fig:eccentricity and orbital tilt}.

We define the stream formation time following \citet{panithanpaisal_galaxy_2021}. 
Figure \ref{fig:disruption distribution} shows the distribution of stream formation lookback times for the full stream populations in \textit{Latte} and \textit{Triple Latte}. Although the distributions overlap substantially, the \textit{Triple Latte} population extends to somewhat earlier formation times, while the \textit{Latte} streams are more concentrated toward recent stream formation times. The distribution also appears bimodal, with one group of streams forming roughly $7-10$ Gyr ago and another forming within the last $\sim 5$ Gyr. This is most likely due to the infall timing for each of the streams, because satellites can begin disrupting into streams around their first pericentric passage after infall. The median stream formation lookback times are $\sim 7.5$ Gyr and $\sim 5.9$ Gyr, respectively. Given the relatively small number of streams in each simulation, however, we interpret this difference as a trend rather than a statistically significant shift. 

\subsection{Pericenters and Apocenters}\label{subsec:peri and apo}

\begin{figure*}[t!] 
    \centering 
    \includegraphics[width=\textwidth]{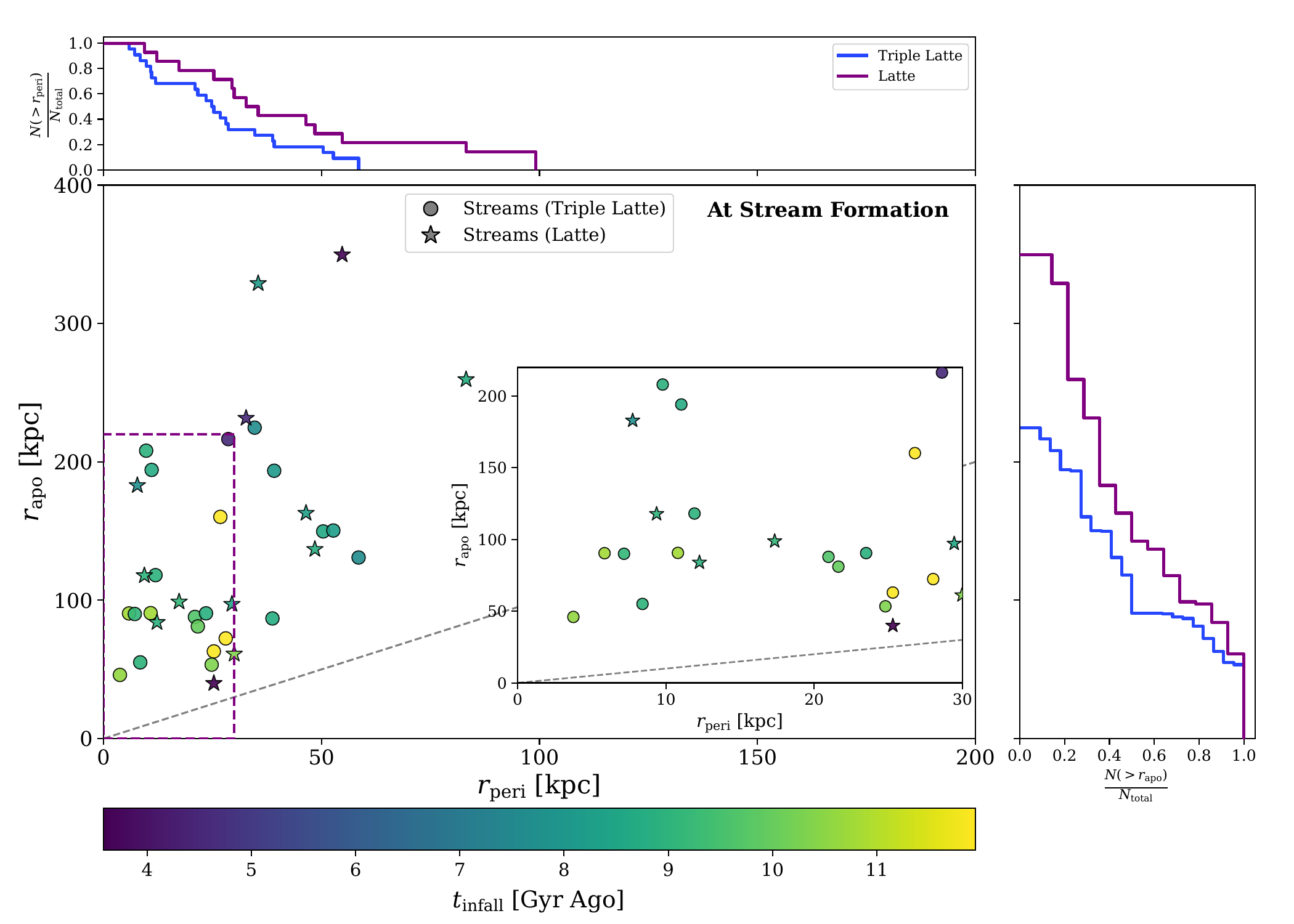} 
    \caption{Pericenters and apocenters of the FIRE-2 \texttt{m12i} streams, measured as the most recent pericentric and apocentric passages nearest prior to stream formation. Streams in \textit{Triple Latte} are denoted with circles, while streams in \textit{Latte} are denoted with stars. The inset zooms in on the subset with $r_{\rm peri} < 30$ kpc and $r_{\rm apo} < 220$ kpc. The top and right panels show the normalized cumulative distributions of pericentric and apocentric distances, respectively, with \textit{Triple Latte} shown in blue and \textit{Latte} in purple. Around the stream formation time, streams in \textit{Triple Latte} tend to have smaller pericentric distances than streams in \textit{Latte}.}
    \label{fig:peri and apo after disruption}
\end{figure*}

\begin{figure*}[t!] 
    \centering 
    \includegraphics[width=\textwidth]{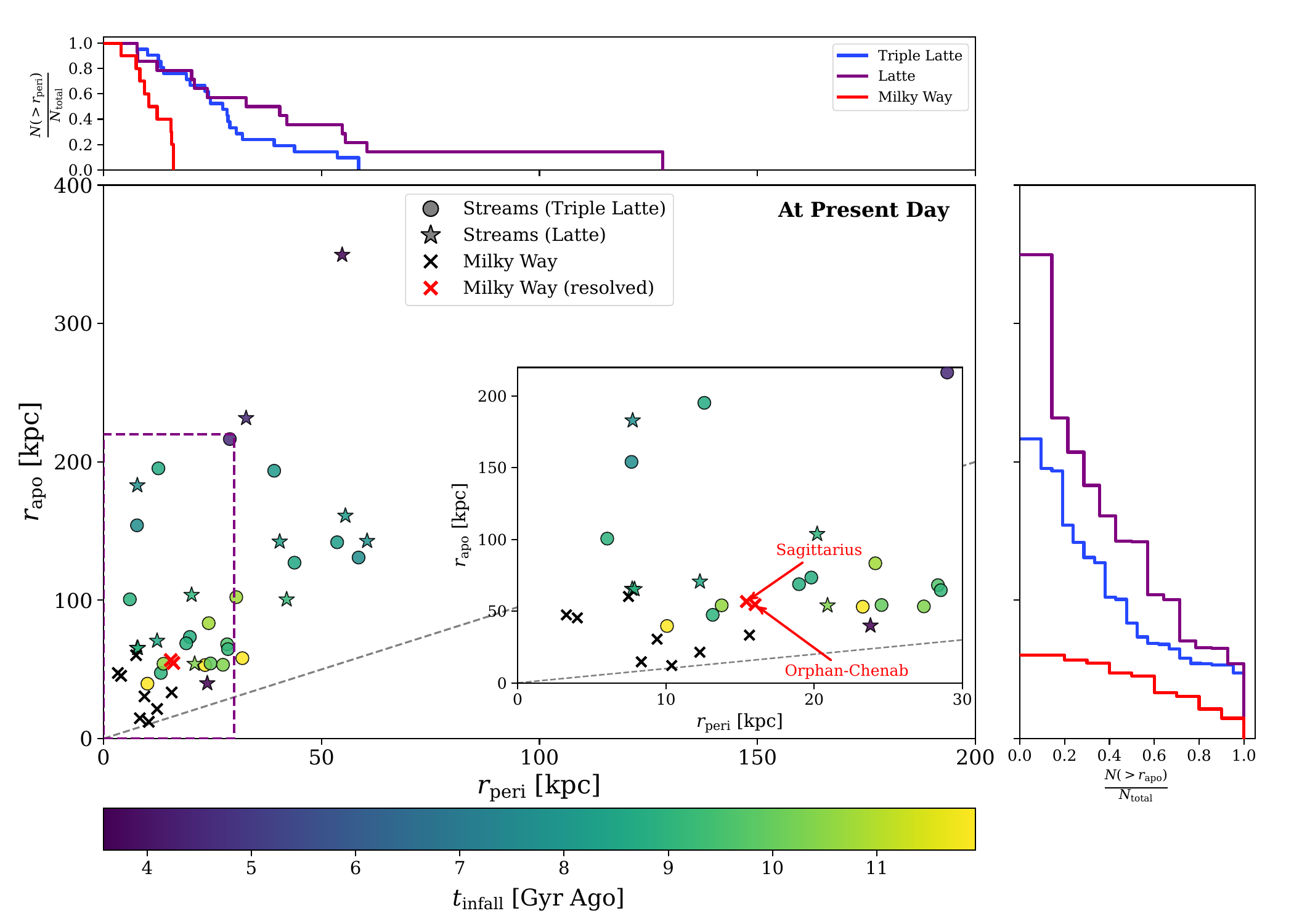} 
    \caption{Most recent pericenters and apocenters of the FIRE-2 \texttt{m12i} and MW streams. For the simulated systems, $r_{\rm peri}$ and $r_{\rm apo}$ correspond to the most recent pericentric and apocentric passages reached by the present day. Crosses indicate MW dwarf galaxy streams, for which we take $r_{\rm peri}$ and $r_{\rm apo}$ from \citet{bonaca_stellar_2024}. We use the same star and circle markers as in Figure \ref{fig:peri and apo after disruption}. The top and right panels show the same normalized cumulative distributions of pericentric and apocentric distances as in Figure \ref{fig:peri and apo after disruption}, with the MW stream distribution added in red. Red crosses indicate MW streams above the stellar-mass threshold resolved by our simulations, while black crosses indicate streams that fall below this threshold based on our selection criteria.
    Two MW streams, Sagittarius and Orphan-Chenab, are labeled in the figure for reference. The circles represent streams at present day in \texttt{m12i} \textit{Triple Latte} while the stars represent streams for \textit{Latte}. The color indicates lookback time to the first infall of the satellite within the virial radius of the host. 
    The inset plot zooms in on a subset of the population that has $r_{\rm peri} < 30$ kpc and $r_{\rm apo} < 220$ kpc. We see that the MW dwarf galaxy stream population coincides with the stream population in \textit{Triple Latte} for our given mass resolution, except for the MW streams shown by black crosses at smaller pericenters ($r_{\rm peri} \lesssim 15 \; \rm kpc$), which fall below the regime resolved by \textit{Triple Latte}.}
    \label{fig:peri and apo}
\end{figure*}


To calculate the pericenters and apocenters of each subhalo, we trace their radial distance from the host across all $600$ output snapshots to reconstruct the subhalo's orbit. Pericenter and apocenter distances change as a function of time due to the time-dependent nature of the galactic potential. Using each subhalo's merger tree, we identify the most recent minimum and maximum distances from the host as the pericenter and apocenter, respectively. Once a pericenter or apocenter is identified, we fit a cubic spline to the surrounding snapshots to more precisely determine the minimum or maximum radial distance from the host. For subhalos that are no longer identified by the merger tree before the present day, we identify their most-bound star particle at the last snapshot in which the subhalo is tracked and follow this particle to the present day. We use the distance of this particle from the host as a function of time to continue tracing the stream's orbit and identify the most recent pericenter and apocenter reached by the present day. \citet{santistevan_modelling_2024} found that the most recent pericenter is typically not the smallest pericenter experienced by a subhalo, indicating that the orbits of most subhalos tend to grow over time. This trend was found across the FIRE-2 MW-mass hosts studied in \citet{santistevan_modelling_2024}. We find a similar trend in our \texttt{m12i} sample, where the majority of streams, intact satellites, and phase-mixed systems in both \textit{Latte} and \textit{Triple Latte} also exhibit increasing pericenter distances over time. This is contrary to the common expectation that satellite orbits generally shrink over time due to dynamical friction \citep[e.g.,][]{binney_galactic_2008}.

Figure \ref{fig:peri and apo after disruption} shows the pericenters and apocenters measured nearest prior to stream formation using the stream formation times calculated in Section \ref{subsec:population statistics}.

Markers are colored by the lookback time at \enquote{accretion}: the time when the progenitor first crosses the virial radius of the host halo; this is also referred to as the \enquote{infall time} in \citet{panithanpaisal_galaxy_2021}. Circles indicate streams in \textit{Triple Latte}, while stars indicate streams in \textit{Latte}. The inset highlights systems from the main panel. The top and right panels show the normalized cumulative distributions for streams in both \textit{Latte} and \textit{Triple Latte}. From the figure, we can see that the streams in \textit{Triple Latte} tend to reach smaller pericentric distances than the streams in \textit{Latte}. This trend is also seen in the cumulative distributions of both the pericenter and apocenter distances, which are shifted toward smaller values. Because these orbital measurements are taken near the onset of stream formation, they provide a more direct comparison of the effects of resolution. The smaller pericenters in the \textit{Triple Latte} streams suggest that higher resolution allows stream progenitors to reach smaller distances from the host before or during disruption, providing a more direct indication that resolution may contribute to the smaller stream pericenters suggested by \citet{shipp_streams_2023}.

Figure \ref{fig:peri and apo} shows the most recent pericenter and apocenter, computed following \citet{shipp_streams_2023}. We use the same marker, color, inset range, and cumulative-distribution conventions as in Figure \ref{fig:peri and apo after disruption}, but additionally include MW stream data here from \citet{bonaca_stellar_2024}.
Crosses show the Milky Way dwarf galaxy stream pericenter and apocenter measurements from \citet{bonaca_stellar_2024}. Red crosses indicate streams above the stellar-mass threshold resolved by our simulations, while black crosses indicate streams that fall below this threshold based on our selection criteria. The red lines show the normalized cumulative distributions of the MW stream pericenter and apocenter distances. 

We find that the streams in \textit{Triple Latte} have smaller recent pericenters and apocenters than streams in \textit{Latte}. 
We find that \textit{Latte} streams have apocenters $r_{apo}>40 \, \mathrm{kpc}$, while we see one stream in \textit{Triple Latte} with $r_{apo}<40 \, \mathrm{kpc}$. We also identify one stream in \textit{Triple Latte} with a pericenter $r_{peri} \lesssim 17 \, \mathrm{kpc}$ while all streams in \textit{Latte} have pericenters $r_{peri} \gtrsim 17 \, \mathrm{kpc}$. Overall, from Figure \ref{fig:peri and apo} we see that \textit{Triple Latte} produces streams with orbits that are broadly consistent with Milky Way observations for streams above the stellar-mass threshold resolved by both simulations (see Section \ref{subsec:peri and apo} for further discussion). However, \textit{Triple Latte} still predicts streams with pericenters $r_{\rm peri} > 50 \; \mathrm{kpc}$ that are not currently observed.

We compare the pericenters and apocenters of stellar streams in the FIRE simulations with those observed in the MW. Figure \ref{fig:peri and apo} shows the pericenters and apocenters of simulated streams together with MW streams. The MW stream orbital properties are taken from \citet{bonaca_stellar_2024}. The cumulative distributions also show that the \textit{Triple Latte} stream population extends to smaller pericenters and apocenters than \textit{Latte}, bringing it into better agreement with the MW distribution. In Figure \ref{fig:peri and apo}, the MW streams occupy a similar region to the stream population identified in \textit{Triple Latte}, particularly at pericenters $\lesssim 70$ kpc where the overlap is most evident in the inset panel of Figure \ref{fig:peri and apo}. Therefore, we find that the MW streams are broadly consistent with the stream population identified in Triple Latte, with the caveat that streams with $r_{\rm peri} \lesssim 15 \; \rm kpc$ are below the mass regime where \textit{Triple Latte} can resolve streams.

As we highlighted in Section \ref{sec:results}, subhalos in the higher-resolution simulation (\textit{Triple Latte}) can remain bound for longer after infall and are generally less disrupted at the present day. Together, these effects allow us to recover pericenters and apocenters closer to the host at higher resolution. 

Previous work by \citet{shipp_streams_2023} found that streams in the \textit{Latte} simulations tended to have larger pericenters than those observed in the MW. In contrast, the stream population in \textit{Triple Latte} shows better agreement with the MW distribution of pericenters and apocenters, suggesting that the higher-resolution simulation captures lower-pericenter stream progenitors that were not resolved in Latte.

However, even in \textit{Triple Latte}, the simulation stream population extends to larger pericenters than those currently observed in the MW. In particular, \textit{Triple Latte} predicts a population of streams with $r_{\rm peri} \gtrsim 50$ kpc that have no observed counterparts. These streams likely have surface brightnesses below current detection limits. Testing this requires mock observations that include observational selection effects \citep[e.g.,][]{shipp_streams_2023}, which we defer to future work.

The current MW stream sample may also be incomplete at larger pericenters, where faint, distant streams are difficult to detect. Roman’s High-Latitude Wide-Area Survey will cover a large area of the Galactic halo and may reveal additional distant stellar streams, potentially extending the observed distribution to larger pericenters \citep{sanderson_near_2026}.

We also label the MW dwarf galaxy streams Sagittarius and Orphan-Chenab for reference in Figure \ref{fig:peri and apo}. These streams, with estimated masses of $\sim 8.91 \times 10^{6} \; \rm M_{\odot}$ \citep{putman_gas_2021} and $\sim 2.0 \times 10^{7} \; \rm M_{\odot}$ \citep{mendelsohn_estimate_2022} respectively, are within the mass ranges where they would be resolved by \textit{Triple Latte}. Because we require subhalos to contain at least $100$ star particles (see Section \ref{subsec:stream candidate identification}), \textit{Triple Latte} can resolve streams with stellar masses $\gtrsim 8.8 \times 10^4$ M$_{\odot}$ initially, corresponding to $\sim 6.2 \times 10^{4}\; \rm M_{\odot}$ at present day due to stellar feedback. Streams with progenitor stellar masses below this threshold ($8.8 \times 10^4$ M$_{\odot}$) are not selected from the simulation and are therefore underrepresented in the simulated sample. Sagittarius and Orphan-Chenab are well within the mass range of what can be resolved in \textit{Triple Latte}, indicating that the higher-resolution simulation reproduces the pericenter and apocenter range of observed MW stellar streams.

As a result, \textit{Triple Latte} should be able to provide a reasonable estimate of the stellar stream populations in MW-like galaxies within this resolved stellar mass regime ($\gtrsim 8.8 \times 10^4$ M$_{\odot}$). 

\section{Matching Across Resolutions}\label{sec:matching}

\begin{figure*}[t!] 
    \centering 
    \includegraphics[width=\textwidth]{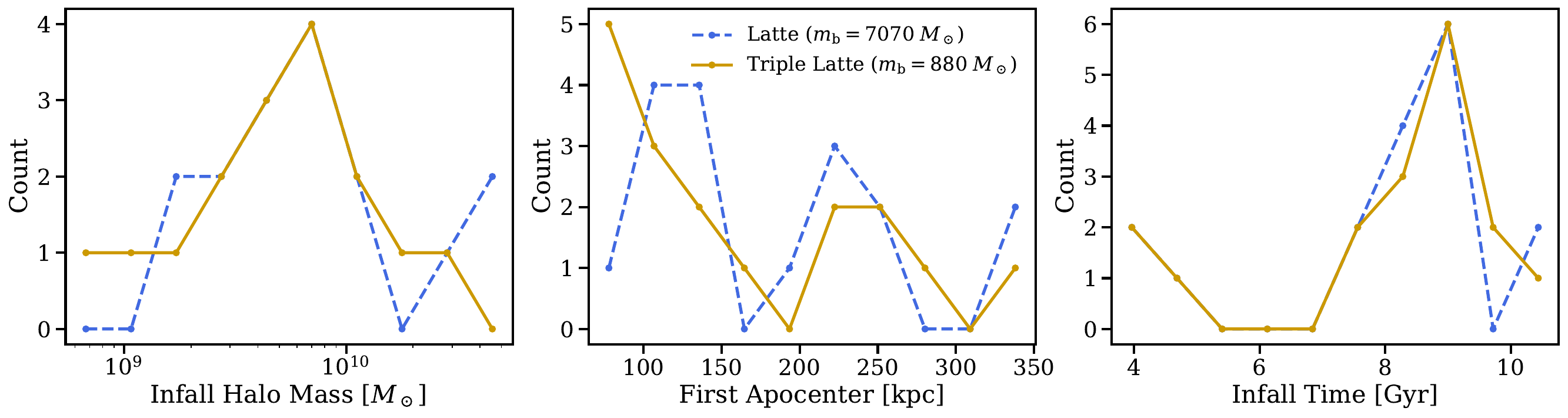} 
    \caption{Distribution plots for \texttt{m12i} \textit{Latte} (blue) and \textit{Triple Latte} (gold) subhalos immediately after first infall. We show the peak mass distribution (left), the first apocenter value after infall (middle), and the infall time (right).}
    \label{fig:infall properties}
\end{figure*}

\begin{figure*}[t!] 
    \centering 
    \includegraphics[width=\textwidth]{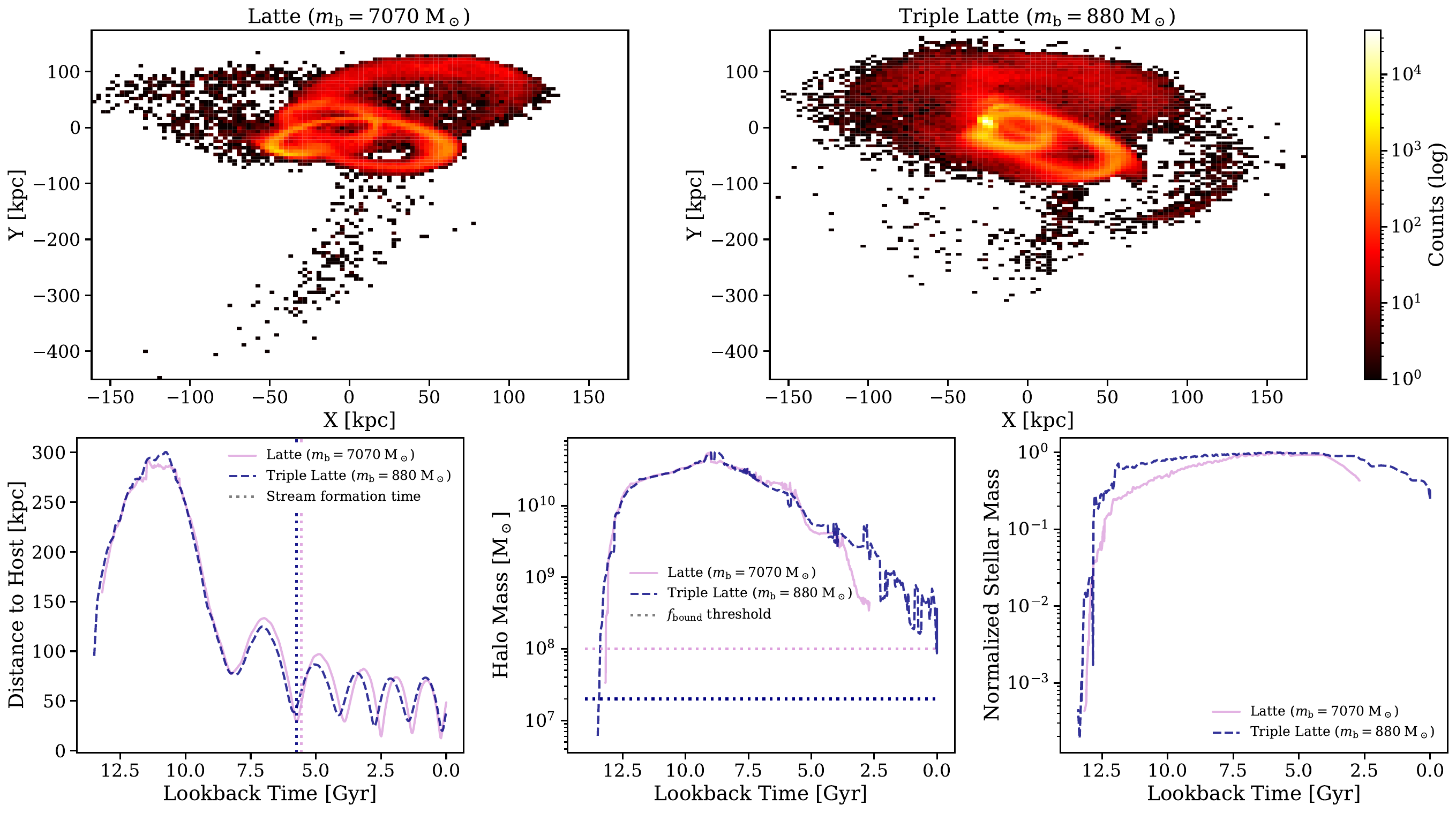} 
    \caption{Example of a matched stream between \texttt{m12i} \textit{Latte} and \textit{Triple Latte}. \textit{Top}: the panels show 2D histograms of star particle positions for both \textit{Latte} (left) and \textit{Triple Latte} (right) at present day. The color bar represents the star particle counts. \textit{Bottom}: orbits (left), halo mass $M_{200\rm m}$ (middle), and normalized stellar mass (right) as a function of lookback time. We define the stellar mass as the sum of the masses of all stars assigned to the subhalo in the halo catalog at each time. The light purple solid line plots these properties for \textit{Latte} and the dark purple dashed line plots the properties for \textit{Triple Latte}. Vertical dotted lines in the bottom-left panel correspond to the times at which the subhalos begin disrupting into streams. Horizontal dotted lines in the bottom middle panel mark the discreteness noise threshold \citep{van_den_bosch_dark_2018}; values above the line are not affected by discreteness noise. Because the $f_{\rm bound}$ threshold depends only on the fixed quantities $N_{\rm acc}$ and $M_{\rm peak}$ for each subhalo, it remains constant with time and is therefore shown as a horizontal line. At higher resolution, we are still able to identify a bound progenitor at present day in \textit{Triple Latte} but not in \textit{Latte}. The two systems begin disrupting at similar times, with \textit{Triple Latte} showing signs of disruption slightly earlier. Despite this, the subsequent evolution differs substantially: around a lookback time of $\sim 3.5-4$ Gyr, \textit{Latte} shows a dramatic drop in halo mass and stellar mass, while \textit{Triple Latte} retains a bound progenitor to the present day.
    }
    \label{fig:match1}
\end{figure*}

\begin{figure*}[t!] 
    \centering 
    \includegraphics[width=\textwidth]{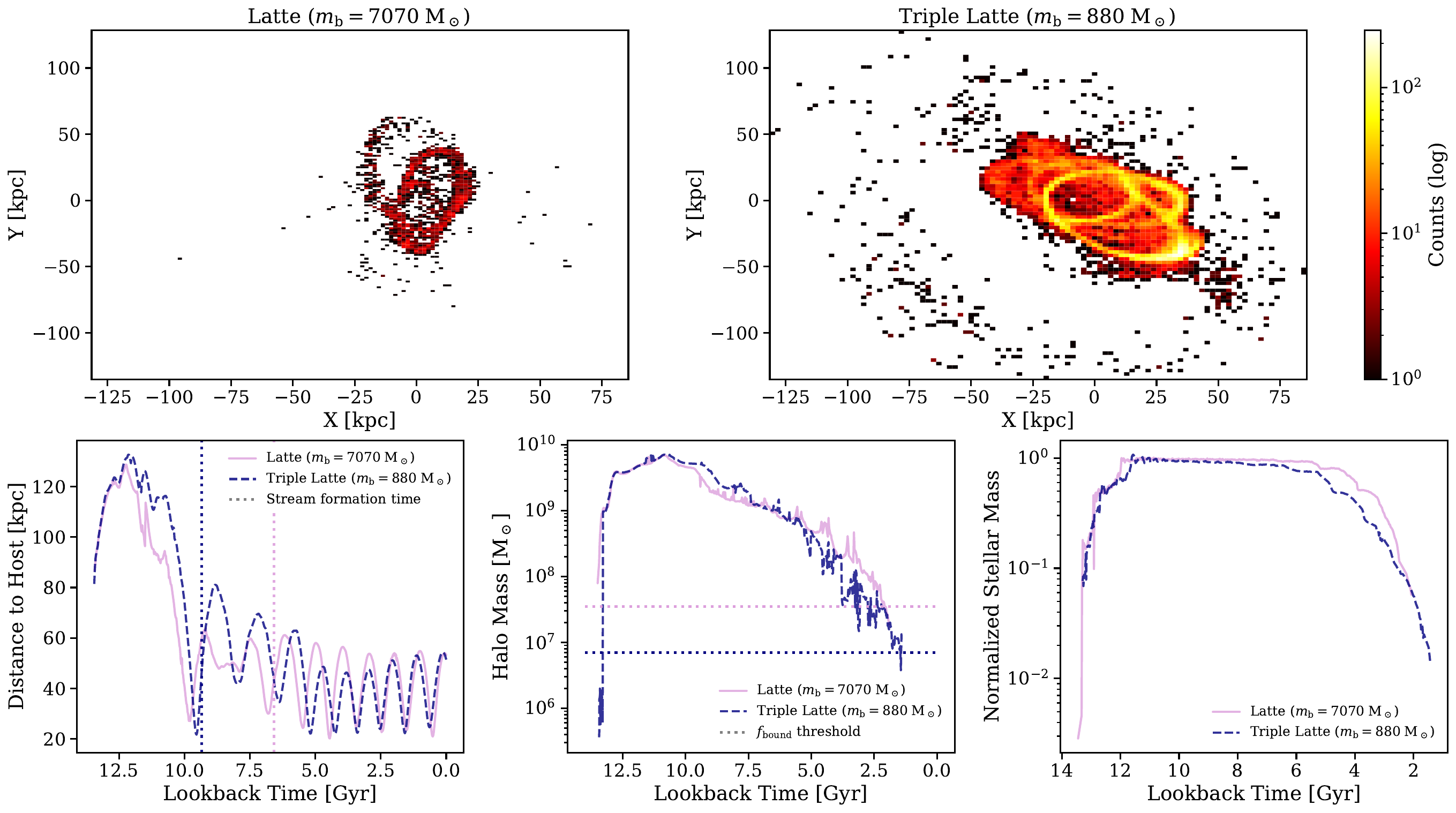} 
    \caption{Similar to Figure \ref{fig:match1}, for a different stream matched between \texttt{m12i} \textit{Latte} and \textit{Triple Latte}. For this matched system, the two simulations show similar orbits at early times, with differences developing after infall. At $z=0$, we find that the stream in \textit{Triple Latte} is more clearly resolved than in \textit{Latte}}
    \label{fig:match2}
\end{figure*}


Cross-matching accretion events between simulations with the same cosmological initial conditions but different resolutions requires an algorithm to match the same structures. For matching stream candidates across resolutions we use star particles associated with their respective subhalos. We use a cross-matching algorithm that is heavily adapted from \citet{riley_auriga_2025}. Unlike \citet{riley_auriga_2025}, who identify matches by tracing dark matter particles back to their positions in the initial conditions, we match subhalos at a fixed early snapshot using their positions and associated star particles.

Our procedure is as follows. First, we locate subhalo positions at $ t \sim 1.81$ Gyr ( $z \sim 3.57$) for both \textit{Latte} and \textit{Triple Latte} stream candidates. We choose $z \sim 3.57$ so that the potential "matched" subhalos are compared at an early time, when their positions in the two simulations are still expected to be similar, but after enough time has passed for star particles to form. These positions correspond to the subhalo centers identified by the merger tree, rather than the individual star particle positions. Using the merger trees, we identify the snapshot at which each subhalo reaches its maximum stellar mass, and then collect the full set of star particle indices at that snapshot in both simulations.
We then compare all possible pairings between candidates in \textit{Latte} and those in \textit{Triple Latte}. For each pairing, we evaluate a set of matching criteria based on the relative subhalo positions across resolutions and consistency in subhalo properties such as orbital trajectories, halo mass, and stellar mass. 

To quantify positional agreement, we compute a distance comparison metric $D$ between all possible pairings between stream candidates, defined as 

\begin{equation}
D = |\vec{\textbf{x}_{H}} - \vec{\textbf{x}_{L}}|,
\end{equation}
where $\vec{\textbf{x}_{H}}$ is the subhalo center position from the \textit{Triple Latte} halo catalog and $\vec{\textbf{x}_{L}}$ is the subhalo center position from the \textit{Latte} halo catalog, both defined in the simulation coordinate frame, which is the same in \textit{Latte} and \textit{Triple Latte}.

We also consider the ratio in star particle number between each stream candidate pair. This is analogous to \citet{riley_auriga_2025}, who use the ratio of dark matter particle numbers for candidate matches across resolutions to check consistency with the resolution scaling. The difference in baryonic mass resolution between \textit{Latte} and \textit{Triple Latte} is a factor of approximately $8.8$. Matched structures that are converged in star particle mass should have a particle ratio of approximately $8.8$. 

To assign a best match for each stream candidate pair in \textit{Latte}, we find the pairing in \textit{Triple Latte} with the smallest value of $D$. In addition, we require that the candidate pair have similar orbits prior to infall. If the pairing has a value $\log{D} < 1.5$ and the particle ratio is between $2$ and $9$, and the pre-infall orbits are similar, the pairing is assigned as a \enquote{match}.

Two out of the $19$ systems in \textit{Latte} do not have a match in \textit{Triple Latte}. This occurs for several reasons related to how stream candidate pairs are compared. Previous studies \citep{riley_auriga_2025} use dark matter particles rather than star particles to perform cross-resolution matching. Since star formation in FIRE is stochastic, matching based on star particles can be less accurate than matching based on dark matter particles. However, we perform the matching using star particles because the merger tree does not have the dark matter particle IDs needed to perform the matching. Another reason for unmatched systems is that, for one of the two unmatched subhalos, the merger tree is unable to track the subhalo over its full orbit. Similar tracking issues may also contribute to some of the orbital differences seen in the matched systems. In addition, as discussed in subsequent paragraphs, matching is more difficult for some subhalos because their orbits can change significantly between the two simulations after infall. In such cases, even if a match at early times is possible, large orbital deviations prevent a match from being identified.


Using the method described, over $89 \%$ ($17$ of $19$) of all stellar stream candidates in \textit{Latte} are matched to their equivalent candidates in \textit{Triple Latte}. These matches include objects that end up as intact satellites, streams, and phase-mixed structures. Two example matches are shown in Figures \ref{fig:match1} and \ref{fig:match2}. We present the case studies of each match in Section \ref{subsec:resolution effects on matched systems}.


The two case studies presented here represent some of the best matches obtained using the matching prescription. However, for most matched systems, the orbits do not continue to closely track each other after infall when comparing between the two resolutions. These orbital differences may reflect small differences between the simulations that grow over time as the systems evolve. This behavior differs from the results presented by \citet{riley_auriga_2025}, who analyze the Auriga simulations and report matching for more than $90 \%$ of subhalos across resolution levels comparable to the resolution increase studied here. The matched streams in \citet{riley_auriga_2025} have very similar morphologies across resolutions, with closer agreement than even the best-case matches presented here.


Many differences between the two simulations may contribute to the orbital differences observed in the matched systems. As discussed in Section \ref{subsec:initial population at infall}, the orientations of the host disks, measured in the same simulation frame, differ by approximately $30^\circ$ at the present day. \citet{santistevan_modelling_2024} found no significant difference in subhalo orbital evolution with disk orientation in idealized tests; however, those tests used a fixed background potential and therefore do not capture the full time-dependent evolution present in cosmological simulations. We also find that the stellar masses of the two hosts differ by approximately $3\%$ at the present day (Section \ref{subsec:initial population at infall}). These differences are not necessarily independent, and other aspects of the evolving hosts may also contribute to the orbital divergence, including differences in the time-dependent gravitational potential, gas distribution, galaxy morphology, star formation and feedback, and interactions with other satellites. Major satellite perturbers, such as an LMC-mass galaxy, could further alter satellite and stream orbits through their tidal influence and their effect on the host halo. Together, these differences may lead to the post-infall orbital differences seen between \textit{Latte} and \textit{Triple Latte}.

\section{Discussion} \label{sec:discussions}

\subsection{Initial Population at Infall}\label{subsec:initial population at infall}


The halo-mass evolution of the central MW of \textit{Latte} and \textit{Triple Latte} remains similar across all redshifts. At $z=0$, the halo and stellar masses differ by $\sim 2\%$ and $\sim 3\%$, respectively. The host galaxies also differ in their present-day disk orientations, with a relative tilt of $\sim 30^\circ$ when measured in the same simulation coordinate frame. We next examine whether the initial populations of infalling subhalos are also similar across the two resolutions. To analyze their properties at infall, we examine the distributions of the systems identified in Section \ref{sec:matching}. The matched subhalos follow broadly similar orbital trajectories at early times, with larger deviations developing primarily after first pericenter. This suggests that the differences seen at later times are not primarily driven by substantially different initial orbital conditions.

We plot in Figure \ref{fig:infall properties} the distributions of the matched stream populations in both \textit{Latte} and \textit{Triple Latte}. The distributions of peak halo mass, first apocenter, and infall time are broadly similar between \textit{Latte} and \textit{Triple Latte}, indicating that the initial stream populations at infall are consistent across resolution. This suggests that orbital differences between matched systems are not primarily driven by substantially different initial orbital conditions. The larger orbital differences appear later, especially after first pericenter, and may reflect differences that develop in the evolving host potential, stellar mass distribution, galaxy structure, and baryonic processes, rather than differences in the initial stream population.

\subsection{Effects of Resolution on Matched Systems}\label{subsec:resolution effects on matched systems}

Differences in the numerical implementation of the Auriga simulations may also contribute to why our matches differ across resolution. Auriga adopts a subgrid interstellar medium (ISM) model with an effective equation of state \citep{grand_auriga_2017}. This model produces converged galaxy properties for systems with stellar masses above $\sim 10^{8} \; \rm M_{\odot}$. The FIRE simulations do not impose an effective equation of state and instead resolve the multiphase ISM directly \citep{hopkins_fire-2_2018}, which leads to increased stochasticity in star formation and can introduce differences in the stellar distributions used to track and match subhalos across resolutions. 

\textbf{Case Study 1:}
Figure~\ref{fig:match1} shows an example of a successfully matched system. The top two 2D histograms show the star particle distributions where the color bar represents star particle counts for both \textit{Latte} (left) and \textit{Triple Latte} (right). In the bottom three subpanels we show the diagnostic properties used for matching, including distance to the host halo (orbital evolution), halo mass, and normalized stellar mass as a function of lookback time, with the different colored lines denoting either \textit{Latte} (light purple) or \textit{Triple Latte} (dark purple). In the bottom-left subpanel of Figure~\ref{fig:match1}, the vertical lines mark the time at which the subhalo begins disrupting into a stream, as described in Section~\ref{subsec:population statistics}. The horizontal lines in the bottom-middle subpanel mark the $f_{\rm bound}$ threshold, which is discussed in more detail below.

From the 2D histograms, we see that at the present day, \textit{Triple Latte} resolves a bound progenitor, whereas \textit{Latte} does not. At a lookback time of $\sim4$ Gyr, the halo mass of the \textit{Latte} system drops sharply, while the \textit{Triple Latte} system retains more of its halo and stellar mass and remains identifiable to the present day. Around the time of this mass loss, the \textit{Latte} system also undergoes a closer pericentric passage than its \textit{Triple Latte} counterpart. Therefore, the different survival histories of this matched pair cannot be attributed to resolution alone, since the post-infall orbital evolution also differs between the two simulations.

The \textit{Triple Latte} system begins disrupting slightly earlier than its \textit{Latte} counterpart. However, despite this earlier onset of disruption, it retains a bound progenitor to the present day, while the \textit{Latte} system becomes fully disrupted. This suggests that the time at which tidal disruption begins does not necessarily determine how quickly the progenitor becomes completely unbound. In this case, the combination of resolution and the differing post-infall orbital histories likely contributes to the different present-day outcomes.

\textbf{Case Study 2:} 

Figure~\ref{fig:match2} shows a second example of a successfully matched system. The subpanels, labels, color notations, and color bars are the same as those in Figure ~\ref{fig:match1}. 

From the 2D histograms, the stream structure is more clearly resolved in the higher-resolution simulation at present day. It is more evident that the stream has wrapped around the host galaxy multiple times and that there is no bound progenitor at the present day, whereas these features are less distinct at lower resolution. Differences in the orbital histories between resolutions become apparent after infall, and the two orbits do not continue to closely track each other. The \textit{Latte} orbit between lookback times of approximately $10$ and $7.5$ Gyr may also be affected by limitations in the halo finder and merger-tree tracking. From the bottom left panel of Figure~\ref{fig:match2}, we see that the \textit{Triple Latte} stream disrupts earlier than its \textit{Latte} counterpart. However, the \textit{Latte} stream appears more morphologically disrupted at present day. As shown in Figure~\ref{fig:disruption distribution}, this behavior is also reflected in the full stream population in both simulations, with \textit{Triple Latte} streams generally disrupting earlier than those in \textit{Latte}.



We also evaluate the numerical disruption criteria from \citet{van_den_bosch_dark_2018}. These criteria were calibrated using idealized dark-matter-only $N$-body simulations with no central galaxy potential, while \textit{Latte} and \textit{Triple Latte} are cosmological baryonic simulations. Therefore, the criteria should be interpreted as guiding diagnostics rather than strict resolution limits for these simulations.

We define the bound mass fraction as

\begin{equation} \label{eq:fbound}
f(\rm t)_{\rm bound} = \frac{M(\rm t)}{M_{\rm peak}}, 
\end{equation}

where $M(t)$ is the dark matter mass of the subhalo at time $t$ and $M_{\rm peak}$ is the maximum dark matter mass reached by the subhalo. We evaluate $f_{\rm bound}$ at the last snapshot in which the subhalo is identified by the merger tree.

Using Equation \ref{eq:fbound}, the first criterion estimates when discreteness noise significantly affects subhalo evolution:

\begin{equation} \label{eq:mass resolution criteria}
f_{\rm bound} > f_{\rm bound, mass} = 0.32(N_{\rm acc}/1000)^{-0.8},
\end{equation}

where $N_{\rm acc}$ is the number of particles in the subhalo at accretion. The second criterion estimates whether the force resolution is sufficient:

\begin{equation} \label{eq:softening resolution criteria}
f_{\rm bound} > f_{\rm bound, force} = \frac{0.79}{f(c)}\left(\frac{\varepsilon}{r_{\rm s,0}}\right)\left(\frac{r_{\rm h}}{r_{\rm s,0}}\right).
\end{equation}

Here, $\varepsilon$ is the gravitational softening length, $r_{\rm s,0}$ is the scale radius at accretion, $r_{\rm h}$ is the half-mass radius, and $f(c) = \ln(1+c) - \frac{c}{1+c}$ is a function of the concentration, $c$.

In the bottom middle panels of Figures \ref{fig:match1} and \ref{fig:match2}, we plot the evolution of the host subhalo mass for each stream. The horizontal lines show the mass scale threshold corresponding to the discreteness-noise criterion in Equation \ref{eq:mass resolution criteria}. The threshold is shown as a horizontal line because $N_{\rm acc}$ and $M_{\rm peak}$ are fixed reference quantities for each subhalo. Subhalo masses above this threshold are not expected to be strongly affected by discreteness noise, while masses below the threshold may be numerically unreliable. Here, $f_{\rm bound}$ is the fraction of the subhalo mass that remains bound relative to its peak subhalo mass, $M_{\rm peak}$. Since Equation \ref{eq:mass resolution criteria} gives a threshold in bound mass fraction, we convert this criterion to a mass threshold by multiplying the right-hand side of Equation \ref{eq:mass resolution criteria} by $M_{\rm peak}$. Therefore, the horizontal line in each panel corresponds to $M_{\rm thresh}=M_{\rm peak}\,0.32(N_{\rm acc}/1000)^{-0.8}$.


We also evaluate the force-resolution criterion in Equation \ref{eq:softening resolution criteria}. The dark-matter force softenings are $\epsilon_{\rm DM}=40 \; \mathrm{pc}$ in \textit{Latte} and $\epsilon_{\rm DM}=20 \; \mathrm{pc}$ in \textit{Triple Latte}. The matched systems do not formally satisfy this criterion (Equation \ref{eq:softening resolution criteria}) in either \textit{Latte} or \textit{Triple Latte}, although they are within an order of magnitude of doing so. This comparison is approximate because the simulations do not directly provide $r_{\rm h}$ for each subhalo. We therefore estimate $r_{\rm h}$ at accretion of each subhalo by assuming an NFW profile with the subhalo virial radius, $R_{\rm vir}$, and scale radius, $r_{\rm s,0}$. 
We use $R_{\rm 200m}$ and $r_{\rm s,0}$ to calculate the concentration, $c=R_{\rm vir}/r_{\rm s,0}$, and infer the enclosed NFW mass profile, $M(<r)/M_{\rm vir}=f(r/r_{\rm s,0})/f(c)$. We then define $r_{\rm h}$ as the radius where $M(<r_{\rm h})=0.5M_{\rm vir}$. This gives an approximate three-dimensional halo half-mass radius for each subhalo, which we use for evaluating the force-softening criterion (Equation \ref{eq:softening resolution criteria}). 


For Match 1 (Figure \ref{fig:match1}), using Equation \ref{eq:fbound} to calculate the bound fraction, the thresholds are $f_{\rm bound} \sim 0.15$ in \textit{Latte} and $\sim 0.070$ in \textit{Triple Latte}. From the halo-mass evolution plot (Figure \ref{fig:match1}: bottom middle), the late-time bound fractions are approximately $f_{\rm bound,force}\sim 0.01$--$0.02$ in \textit{Latte} and $f_{\rm bound,force}\sim 0.008$--$0.015$ in \textit{Triple Latte}. Thus, both the \textit{Latte} and \textit{Triple Latte} systems fall below the formal force-resolution threshold at late times ($f_{\rm bound} > f_{\rm bound, force}$). For Match 2 (Figure \ref{fig:match2}), the estimated force-resolution thresholds are $f_{\rm bound,force}\sim 70$ in \textit{Latte} and $f_{\rm bound,force}\sim 0.45$ in \textit{Triple Latte}. The late-time bound fractions are approximately $f_{\rm bound,force}\sim 0.001$--$0.002$ in both simulations, placing both systems below the formal force-resolution threshold.

Although Equation \ref{eq:softening resolution criteria} is not formally satisfied, this result should be interpreted with caution. The criteria from \citet{van_den_bosch_dark_2018} were derived from controlled dark-matter-only simulations and may not fully capture the behavior of subhalos in a cosmological baryonic environment. Most basically, the inclusion of a central galaxy potential can alter the disruption of subhalos relative to idealized dark-matter-only experiments. More specifically, effects such as disk shocking, baryonic contraction, time-dependent tides, and differences in the host potential can alter subhalo disruption relative to idealized dark-matter-only experiments \citep{shipp_auriga_2025}. There are currently no equivalent resolution criteria calibrated for simulations with baryons, so it is unclear how the thresholds from \citet{van_den_bosch_dark_2018} would change. It is also unclear whether including baryons would make these criteria more or less restrictive. Therefore, these criteria provide useful resolution diagnostics, but they should not be treated as definitive disruption thresholds for \textit{Latte} and \textit{Triple Latte}.

\textbf{Key takeaways:} 

In both case studies, the \textit{Triple Latte} progenitors stay bound for longer, even when stream formation starts at a similar or slightly earlier time than in \textit{Latte}. This means higher resolution can help the progenitor survive longer after disruption begins. Even though the matched orbits differ after infall, the fact that the \textit{Triple Latte} streams begin disrupting at similar or earlier times while their progenitors remain bound for longer suggests that resolution could contribute to their longer survival. 

The orbits also change differently between the two simulations after infall. Because of this, we cannot say that resolution alone causes the different disruption outcomes. The numerical disruption criteria \citep{van_den_bosch_dark_2018} show that some systems may be affected by numerical effects, but these criteria were developed for idealized dark-matter-only simulations and should be treated with caution here.

\subsection{Catalog Completeness} \label{subsec:completeness}

\begin{figure*}[t!] 
    \centering 
    \includegraphics[width=0.85\textwidth]{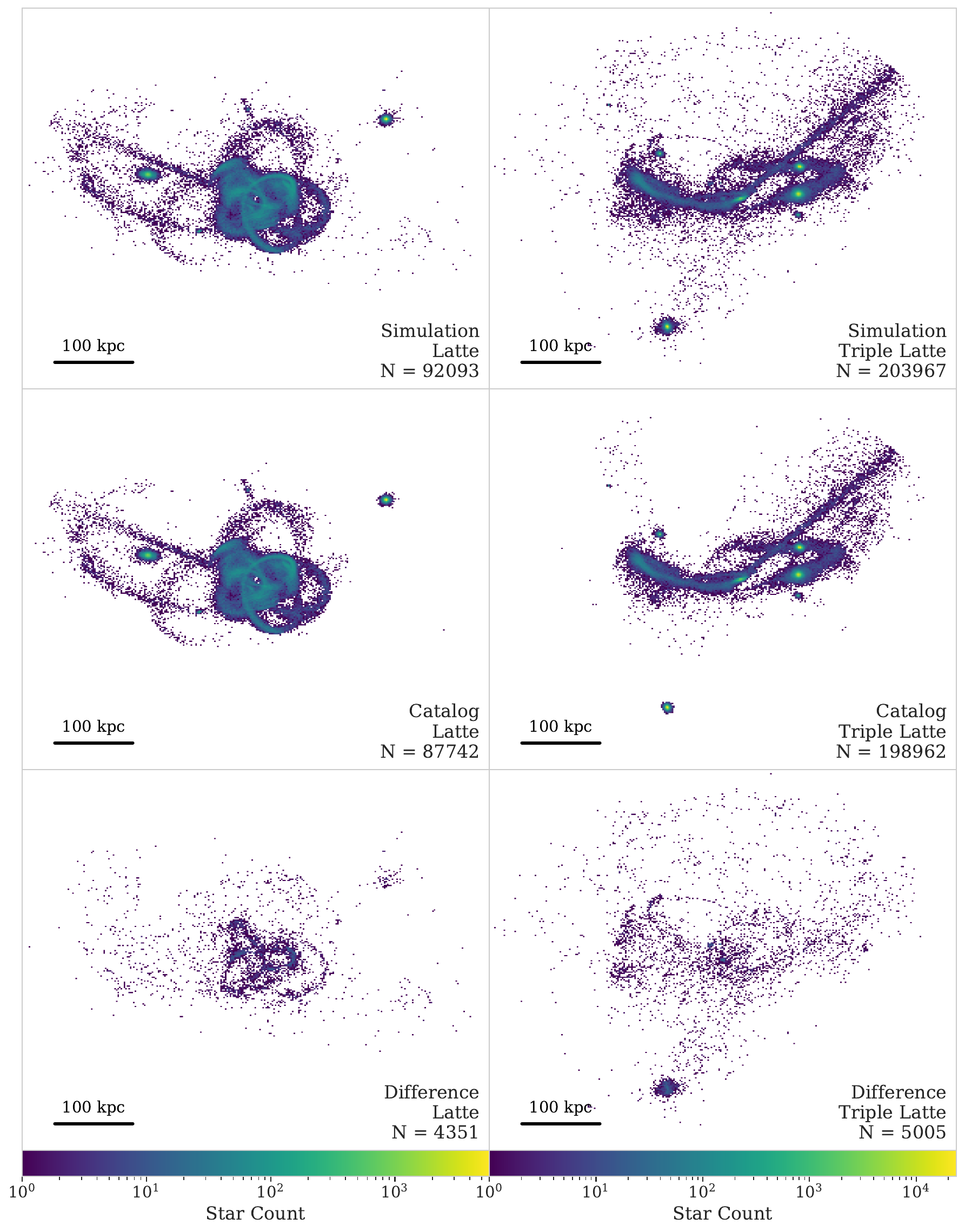} 
    \caption{Two-dimensional histogram plots analyze our catalog completeness in the Latte and Triple Latte simulations. The top, middle, and bottom panels show the simulation predictions, catalog data, and their differences, respectively. The left and right columns present results for \textit{Latte} and \textit{Triple Latte}. We apply several selection criteria to isolate relevant star particles. First, we restrict the star formation distance to between $30$ and $550 \; \rm kpc$ to exclude stars from the host galaxy and distant field stars. Second, we require a formation time greater than $6 \; \rm Gyr$, consistent with our stream selection criteria (Section \ref{sec:stream selection}) that identifies subhalos at $z \lesssim 1$. Finally, we limit star particles to those within $300 \; \rm kpc$ of the host, ensuring they remain within the host's virial radius.}
    \label{fig:completeness}
\end{figure*}
A key consideration in constructing stream catalogs for both \textit{Latte} and \textit{Triple Latte} is completeness. Here, completeness is defined as whether the identification process recovers all streams and their associated star particles in the simulation.

As discussed in Section \ref{subsec:stream candidate identification}, we identified bound subhalos at five different redshifts between $z\sim 0.0-1.0$ to ensure that all stream candidates are identified in the simulations. This approach allows us to identify subhalos that may not be identified by the merger tree at the present day, particularly those that have undergone complete disruption. The upper limit $z \sim 1.0$ was chosen because the subhalos that become unbound before $z \sim 1$ have all become phase-mixed by the present day.


To assess catalog completeness, we select a region of star-particle parameter space where we expect the catalog to recover the relevant accreted stars. We then compare the star particles in the simulation to those included in the catalog. Stars that satisfy the same selection but are missing from the catalog are counted as incomplete. We select stars based on their formation time, formation distance from the host, and present-day distance from the host. These selection criteria exclude host and field stars and limit the comparison to the range covered by our catalog. Systems that became unbound before $z\sim1$ are not expected to be included, since they may already be phase-mixed.

Figure \ref{fig:completeness} shows the results. The left and right columns correspond to Latte and Triple Latte respectively. The top panels show star particles directly from the Latte and Triple Latte simulations meeting our criteria. The middle panels show the star particles identified in our catalog using the same selection criteria. The bottom panels display the particles in the simulation but missing from our catalog.

We find completeness of $95.3\%$ for \textit{Latte} and $97.4\%$ for \textit{Triple Latte}. The missing $\sim 3$--$5\%$ of particles likely stems from two sources. First, the underlying halo catalog for each simulation may not identify all star particles associated with a given subhalo. One possible reason is because the halo catalogs are constructed from dark matter particles, and star particles are assigned to subhalos only during post-processing \citep{wetzel_public_2023}, which can introduce mismatches depending on how that assignment is performed.

Additionally, we sample subhalos at $5$-snapshot intervals (Section \ref{subsec:stream candidate identification}) to balance completeness and purity. Sampling every snapshot would recover more star particles but risk introducing contamination from the host halo. 

\section{Conclusion} \label{sec:conclusions}

In this paper, we compare stellar stream populations in the FIRE-2 simulations of the same galaxy \texttt{m12i}, both at the resolution of the \textit{Latte} suite ($m_{\rm baryon} = 7070 M_\odot$) and at $8 \times$ higher mass resolution in the \textit{Triple Latte} suite ($m_{\rm baryon} = 880 M_\odot$), to understand how numerical resolution affects the identification, evolution, and orbital properties of disrupted satellites. The higher mass resolution in \textit{Triple Latte} leads to a larger sample of stream candidates at lower stellar masses. The important findings are the following:

\begin{enumerate}
\item We find that the pericenter–apocenter distribution of streams in \textit{Triple Latte} is broadly consistent with that of MW stellar streams. This suggests that higher-resolution simulations are better able to capture the orbital properties of stream progenitors observed in the MW, particularly for systems with smaller pericenters. Within the resolved stellar mass regime, \textit{Triple Latte} therefore provides a more complete picture of the stellar stream population in MW-mass galaxies.

\item The higher resolution in \textit{Triple Latte} allows disrupted subhalos to reach smaller galactocentric radii (see Figure \ref{fig:radial distribution}). In \textit{Latte}, some systems may disrupt more rapidly because the lower mass resolution makes them more susceptible to numerical effects. In \textit{Triple Latte}, these same types of systems can remain identifiable for longer after infall, allowing their orbits to carry them closer to the host galaxy while retaining a bound progenitor. This helps explain why Triple Latte better samples streams with smaller pericenters and shows better agreement with the observed MW stream population.

\item Near the time of stream formation, the \textit{Triple Latte} streams tend to have smaller pericenters than the \textit{Latte} streams. Because these orbital measurements are taken close to the onset of stream formation, they provide a more direct indication that resolution may contribute to bringing the streams closer in at the time of stream formation.


\item The orbital evolution of infalling satellites can differ substantially between simulations, even when the hosts have similar final masses and broadly similar mass-assembly histories. The matched systems develop orbital differences after infall and do not always continue to closely track each other. Such differences can arise from small numerical perturbations and the intrinsically stochastic evolution of the simulations. This indicates that their subsequent evolution is sensitive to small differences that develop and grow over time between the two simulations. These may include differences in the evolving host potential and central galaxy structure.


\item Although \textit{Triple Latte} improves agreement with the observed MW stream population, the disrupted subhalo population may still remain sensitive to numerical resolution. Convergence tests with even higher-resolution simulations are needed to determine whether the disruption of low-mass subhalos and the resulting stellar stream population are fully resolved.
\end{enumerate}

Overall, this work shows that increasing numerical resolution improves the orbital characterization of stellar streams in the case of the FIRE-2 \textit{Latte} and \textit{Triple Latte} simulations. In \textit{Triple Latte}, higher resolution reveals more low-mass streams, allows disrupted systems to remain identifiable for longer, and better recovers streams on orbits that pass closer to the host galaxy (finding 2). However, this disrupted subhalo population is still not fully resolved, and the differences between matched systems show that resolution is not the only factor that shapes the final stream population. Even for the same initial conditions and similar host masses, differences in the orientation and evolution of the central disk can strongly affect satellite orbits after infall (finding 4). Quantifying these resolution effects is therefore important for separating physical effects from numerical limitations. Future simulations with higher resolution along with mock observations will therefore be needed to fully understand the complete stellar stream populations expected around MW-mass galaxies.



\begin{acknowledgments}
We thank Ana Bonaca, Alexander Riley, and Nora Shipp for useful discussions. 
Numerical calculations and plots in this paper were run on the Texas Advanced Computing Center (TACC) allocation AST21010.
Support for P.F.H. was provided by a Simons Investigator grant.
A.W. received support from NSF CAREER award AST-2045928.
\end{acknowledgments}

\appendix

\software{astropy \citep{the_astropy_collaboration_astropy_2013, the_astropy_collaboration_astropy_2018, the_astropy_collaboration_astropy_2022}, Matplotlib \citep{hunter_matplotlib_2007},
NumPy \citep{harris_array_2020}, \texttt{consistent-trees} \citep{behroozi_gravitationally_2013}, \texttt{rockstar} \citep{behroozi_rockstar_2013}, \texttt{halo\_analysis} \citep{wetzel_haloanalysis_2020}, \texttt{gizmo\_analysis} \citep{wetzel_gizmoanalysis_2020}}



\bibliography{references}{}
\bibliographystyle{aasjournal}



\end{document}